\documentclass{aa}  

\def\heao1{{\it HEAO-1\/}}

\def\lsimeq{{_<\atop^{\sim}}}
\def\lesssim{\mathrel{\hbox{\rlap{\hbox{\lower4pt\hbox{$\sim$}}}\hbox{$<$}}}}
\def\gtrsim{\mathrel{\hbox{\rlap{\hbox{\lower4pt\hbox{$\sim$}}}\hbox{$>$}}}}

\usepackage{hyperref}	
\hypersetup{colorlinks=true,linkcolor=blue,citecolor=blue,filecolor=blue,urlcolor=blue}
\usepackage{txfonts}
\usepackage{rotating}

\begin{document}

\title{The ALPINE-ALMA [CII] survey: Dust mass budget in the early Universe }

\author{F. Pozzi\inst{1,2}, F. Calura\inst{2}, Y. Fudamoto\inst{3},
  M. Dessauges-Zavadsky\inst{3}, C. Gruppioni\inst{2},
  M. Talia\inst{1}, G. Zamorani\inst{2}, M. Bethermin\inst{4},
  A. Cimatti\inst{1,5}, A. Enia\inst{1}, Y. Khusanova\inst{4}, R. Decarli\inst{2}, O. Le F\`{e}vre\inst{4},
  P. Capak\inst{6}, P. Cassata\inst{7}, A.L. Faisst\inst{6}, L. Yan\inst{8},
  D. Schaerer\inst{3}, J. Silverman\inst{9}, S. Bardelli\inst{2},
  M. Boquien\inst{10}, A. Enia\inst{1}, D.Narayanan\inst{11}, M. Ginolfi\inst{12}, N.P. Hathi\inst{13},
  G.C. Jones\inst{14,15}, A.M. Koekemoer\inst{13},
  B.C. Lemaux\inst{16}, F. Loiacono\inst{1},
  R. Maiolino\inst{14,15,17}, D.A. Riechers\inst{18},
  G. Rodighiero\inst{6}, M. Romano\inst{7,19}, L. Vallini\inst{20},
  D. Vergani\inst{2}, E. Zucca\inst{2}
}

{
\institute{Dipartimento di Fisica e Astronomia, Università of Bologna, via Gobetti 93/2, 40129, Bologna, Italy
\and
INAF -- Osservatorio di Astrofisica e Scienza dello Spazio di
Bologna, via Gobetti 93/3, 40129, Bologna, Italy
\and
Department of Astronomy, University of Geneva, 51 Ch. des
Maillettes, 1290 Versoix, Switzerland
\and
Aix Marseille Univ. CNRS, LAM, Laboratoire d'Astrophysique de
Marseille, Marseille, France
\and
INAF -- Osservatorio astrofisico di Arcetri, Largo E. Fermi 5,
I-50125, Firenze, Italy
\and
Infrared Processing and Analysis Center, California Institute of
Technology, Pasadena, CA 91125, USA
\and
Dipartimento di Fisica e Astronomia, Università di Padova, vicolo
Osservatorio 3, 35122 Padova, Italy
\and
The Caltech Optical Observatories, California Institute of Technology,
Pasadena, CA 91125, USA
\and
Kavli Institute for the Physics and Mathematics of the Universe, The University of Tokyo, 5-1-5 Kashiwanoha, Kashiwa-shi,
Chiba, 277-8583 Japan
\and
Centro de Astronomia (CITEVA), Universidad de Antofagasta, Avenida
Angamos 601, Antofagasta, Chile
\and
Department of Astronomy, University of Florida, 211 Bryant Space Sciences Center, Gainesville, FL 32611 USA
\and
European Southern Observatory, Karl-Schwarzschild-Straße 2, 85748
Garching, Germany
\and
Space Telescope Science Institute, 3700 San Martin Drive, Baltimore,
MD 21218, USA
\and
 Cavendish Laboratory, University of Cambridge, 19 J. J. Thomson Ave.,
 Cambridge CB3 0HE, UK
 \and
16 Kavli Institute for Cosmology, University of Cambridge, Madingley
Road, Cambridge CB3 0HA, UK
\and
Department of Physics, University of California, Davis, One Shields
Ave., Davis, CA 95616, USA
\and
Department of Physics and Astronomy, University College London, UK
\and
Department of Astronomy, Cornell University, Space Sciences Buildin,
Ithaca, NY 14853, USA
\and
INAF -- Osservatorio Astronomico di Padova, Vicolo dell'Osservatorio 5, I-35122, Padova, Italy
\and
Scuola Normale Superiore, Piazza dei Cavalieri 7, I-56126 Pisa, Italy
}

\authorrunning{F. Pozzi, F. Calura, Y. Fudamoto et al.}
\titlerunning{The ALPINE-ALMA survey: Dust mass budget in the early Universe}

\date{Accepted May 26, 2021}


\abstract
{}
{The dust content of normal galaxies and the dust mass density (DMD)
  at high-$z$ ($z>4$) are unconstrained given the source confusion
  and the sensitivity limitations of previous observations. The ALMA Large Program to INvestigate [CII] at
  Early Times (ALPINE), which targeted 118 UV-selected 
  star-forming galaxies at 4.4$<$z$<$5.9, provides a new opportunity
  to tackle this issue for the first time with a statistically robust dataset. 
}
{We have exploited the rest-frame far-infrared (FIR) fluxes of the 23 continuum
  individually detected galaxies and stacks of continuum images to measure the dust
  content of the 118 UV-selected ALPINE galaxies. We have focused on  
  the dust scaling relations and, by
  comparing them with predictions from
  chemical evolution models, we have probed the evolutionary
  stage of UV-selected galaxies at high-$z$.  By using the observed correlation
  between the UV-luminosity and the dust mass, we have estimated the DMD
  of UV-selected galaxies at $z{\sim}5$, weighting the galaxies by means of the UV-luminosity
function (UVLF). The derived DMD has been compared with the value we
have estimated from the 10 ALPINE far-IR continuum blindly detected galaxies at the redshift of the
ALPINE targets.}
{Our ALMA survey allows the exploration for the first time of the dust content
in normal star-forming galaxies at $z>4$ in a statistically robust
sample of sources. The comparison of the observed dust scaling
relations with chemical evolution models suggests that ALPINE galaxies 
are not likely progenitors of disc galaxies, but of 
intermediate and low mass proto-spheroids, resulting in present-day bulges
of spiral or elliptical galaxies.
Interestingly, this conclusion is in line with the independent morphological
analysis, that shows that the majority (${\sim}$70\%) of the dust-continuum detected galaxies have a disturbed morphology. 
The DMD obtained at $z{\sim}5$ from UV-selected sources is
$\sim$30~$\%$ of the value obtained from blind far-IR selected sources, showing that the UV-selection
misses the most dust-rich, UV-obscured galaxies.}

{}

\maketitle

%

\section{Introduction}
\label{intro_sec}

Cosmic dust accounts for an almost negligible contribution of the baryon mass in the
Universe ($\sim$0.1 \% in the local Universe, see \citealt{2012ApJ...759...23S}). Nevertheless, it plays a crucial  
role in many astrophysical and astrochemical aspects. First among others, it strongly
affects the spectral energy distribution of galaxies,
being a source of attenuation for UV/optical photons and an emission
source in the infrared domain. Therefore, it is of primary importance
to recover the galaxy dust properties to achieve a self-consistent
understanding of the physics and evolution of galaxies across the
cosmic time. 

Before the advent of the Atacama Large Millimeter Array (ALMA), dust
emission of normal
star-forming galaxies was detected mainly thanks to the Herschel
surveys up to $z<4$ ($z{\sim}$2,
e.g., \citealt{2011ApJ...739L..40R}; \citealt{2013MNRAS.432...23G};
\citealt{2013A&A...553A.132M}; \citealt{2014A&A...572A..90L}) or extreme dusty (usually
strongly lensed) galaxies up to 
high-$z$ ($z>4$, e.g., \citealt{2012A&A...538L...4C}, \citealt{2016MNRAS.461.1100R}, \citealt{2017MNRAS.465.3558N}).

Nowadays, ALMA is revolutionising this field of research thanks to its
superb sensitivity and high-spatial resolution that avoids the problem of source confusion, allowing
continuum detection at $z>4$  also for
normal star-forming galaxies (i.e. with star-formation rate from a few up to tens of solar masses/year).

Given its relatively small field of
view ($\sim$30~arcsec at 1.2~mm), ALMA started the exploration of the high-$z$ Universe by targeting
individual galaxies (e.g \citealt{2013ApJ...768...91H}; \citealt{2014ApJ...796...84R},
\citealt{2015Natur.522..455C}, \citealt{2015Natur.519..327W};
\citealt{2017ApJ...837L..21L}). The number of normal/large programs
over the last years has been steadily increasing
and different continuum surveys have been performed, from the deepest
 observations over small areas ($<$5 arcmin$^2$, e.g.,
 \citealt{2017MNRAS.466..861D}, \citealt{2020ApJ...897...91G})
 to the wider (few tens of arcmin$^2$) and shallower surveys (e.g., \citealt{2018A&A...620A.152F}).

\cite{2020ApJ...892...66M} carried out the first study of the dust mass
density (DMD) from $z{\sim}$0.5 up to $z{\sim}$5, taking advantage of the 
deepest (9.5~$\mu$Jy/beam) ALMA 1.2mm continuum map of $\sim$4
arcmin${^2}$ in the Hubble Ultra Deep Field (HUDF) of the survey ASPECS (\citealt{2016ApJ...833...67W}, \citealt{2019ApJ...882..138D}, \citealt{2020ApJ...897...91G}). Their results confirm the presence of a peak around $z{\sim}$1-3 and a decrease of the DMD from $z{\sim}$1 down to the local Universe already
found by Herschel (\citealt{2011MNRAS.417.1510D}; 
\citealt{2018MNRAS.475.2891D}, \citealt{2020MNRAS.491.5073P}), the latter result being 
 in contrast with cosmological simulations (e.g., \citealt{2017MNRAS.471.3152P},
\citealt{2018MNRAS.478.4905A}, \citealt{2019MNRAS.490.1425L}).

In order to improve our understanding of the DMD at high-$z$ (>4), in
this work we take advantage of our recently completed ALMA Large program to
INvestigate [CII] at Early times (ALPINE, PI: Le F\`{e}vre, see \citealt{2020A&A...643A...1L}, \citealt{2020A&A...643A...2B},
\citealt{2020ApJS..247...61F}). 
The goal of the ALPINE survey was to observe the prominent  [CII]
  158~$\mu$m emission line for 118 UV-selected normal star-forming
  galaxies at $z{\sim}4.4-5.8$.  From 118 targets, 75 galaxies were
detected in [CII] and 23 in continuum. Besides the main targets, also a blind search
for continuum and line emitters in an ${\sim}25$ arcmin$^{2}$ area has been
performed in the ALPINE pointings (\citealt{2020A&A...643A...2B},
\citealt{2020arXiv200604837L}). 

 Over the last year, many works based on the ALPINE survey have been presented, describing
the Interstellar Medium (ISM) properties of `normal'
star-forming galaxies at $z{\sim}$5 (the [CII]-SFR
relation: \citealt{2020arXiv200200979S};
the [CII] outflows: \citealt{2020A&A...633A..90G}; 
the [CII] spatial scales: \citealt{2020arXiv200300013F};
the Ly$\alpha$-[CII] velocity offset: \citealt{2020A&A...643A...6C};
the IRX-$\beta$ relation: \citealt{2020arXiv200410760F};
 the gas content: \citealt{2020A&A...643A...5D}, detailed studies on individual
sources (\citealt{2020MNRAS.491L..18J}; \citealt{2020MNRAS.496..875R};
 \citealt{2020A&A...643A...7G}) or
presenting statistical studies at $z{\sim}$5, including the [CII]
luminosity function for ultra-violet (UV) and serendipitous detected
line emitters (\citealt{2020arXiv200604835Y}, \citealt{2020arXiv200604837L}),
the infrared (IR) luminosity function (\citealt{2020A&A...643A...8G}) and the cosmic star formation rate density (SFRD, \citealt{2020arXiv200708384K}).
 
In the present work, we measure the dust content of ~UV-selected
`normal' (main-sequence) galaxies and IR continuum serendipitously
detected sources. In Sect. \ref{data_sec} and \ref{dust_sec} the ALPINE
sample and the measurements of the dust masses are presented, respectively.
In Sect. \ref{physics_sec}, the derived ISM properties of the
ALPINE targets (dust masses from the
present work and gas masses from \citealt{2020A&A...643A...5D}) are
combined with the well characterised stellar masses and SFRs
(\citealt{2020ApJS..247...61F}), to discuss the dust scaling
relation for the UV-population at $z{\sim}5$. 
In Sect. \ref{dmd_sec} the DMD for the
UV-selected and for the IR-selected populations are presented. Finally, in
Sect. \ref{summary_sec} we present our conclusions.

Throughout the paper, we assume a $\Lambda$CDM cosmology with
$\Omega_{m}=0.3$, $\Omega_{\lambda}=0.7$ and $H_{0}$=70
km~s$^{-1}$Mpc$^{-3}$.

\section{Observations}
\label{data_sec}

In the following section we describe in detail the ALPINE continuum detected
sample that we use in our study. For a more complete description of
the overall survey, of the data reduction, and of the ancillary
data we refer to \cite{2020A&A...643A...1L}, \cite{2020A&A...643A...2B} and \cite{2020ApJS..247...61F}, respectively.

The 118 galaxies from the ALPINE survey (\citealt{2020A&A...643A...1L}) are rest-frame UV-selected
galaxies at redshift $z{\sim}4.4-5.9$. These galaxies originated in two
fields, namely the `COSMOS Evolution Survey' field (COSMOS,
\citealt{2007ApJS..172....1S}) and the `Extended Chandra Deep Field
South' field (ECDFS, \citealt{2002ApJS..139..369G}). They all have spectroscopic redshifts from different campaigns (see
\citealt{2020ApJS..247...61F} for details). They were selected to be
representative of star-forming Main-Sequence Galaxies at redshift
$z{\sim}$5 (i.e. \citealt{2014ApJS..214...15S}). They have stellar masses in the range M$_{\star}=10^{8.4}-10^{11}$ M$_{\odot}$ and star-formation rate in
the range SFR=3-270 M$_{\odot}$yr$^{-1}$. The stellar masses are
derived from SED-fitting using broadband and medium
(intermediate) band rest-frame UV/optical photometry (see
\citealt{2020ApJS..247...61F} for details). The SFRs reported are
derived from \cite{2020arXiv200200979S}, using the ALPINE far-IR data,
to properly account for dust-obscured SFR. They have been computed as
SFR(UV)+SFR(IR) in case of IR continuum detection, and are derived in
the other cases from the observed UV slope and luminosity using the
ALPINE IRX-beta relation (obtained from stacking, see \citealt{2020arXiv200410760F}).
These SFRs are globally consistent with the SFR derived from
UV/optical SED-fitting
(median(SFR(UV+IR)/SFR(SED))${\sim}$1.2). Anyway, since there are few
outliers for which SFR(UV+IR) results to be a factor up to ${\sim}5$ higher
than SFR(SED), we prefer to consider the SFR(UV+IR) in the present
analysis to properly account for dust-obscured SFR.

The ALPINE targets were observed during Cycle 5 and Cycle 6 in Band 7
(275-373 GHz) which covers the [CII] line from $z=4.1-5.9$. To avoid
atmospheric absorption, no targets were included  in the redshift range
$z=4.6-5.1$. Each target was observed between 15 and 45 minutes. The
data were reduced and calibrated with standard routines using the
`Common Astronomy Software Application' (CASA) software
(\citealt{2007ASPC..376..127M}).  We refer to
\cite{2020A&A...643A...2B} for all the details of the data reduction processes and the
catalogue construction. The continuum maps - that are of
interest for this work - were produced using line free
channels, reaching an angular resolution between 0.7$^{\prime\prime}$ and
1.6$^{\prime\prime}$ and mean sensitivities (RMS of 50 $\mu$Jy/beam and 28
$\mu$Jy/beam in the 4.3$<z<4.6$ and 5.1$<z<5.9$ ranges,
respectively. Some sources were marginally resolved at the angular
resolution reached by the observations and different methods were implemented and compared to
measure the fluxes. The simulations performed (see Sect. 3.2 and 3.9 in \citealt{2020A&A...643A...2B}) showed
that the 2D-fit photometry is the most accurate one, even though other
methods are consistent within the error bars. The final continuum catalogue consists of 23 sources detected at
signal-to-noise ratio SNR$>$ 3.5 (see Table B.1 in \citealt{2020A&A...643A...2B}). In case of
multi-components objects ($DEIMOS\_COSMOS\_881725$, $vuds\_cosmos\_5101209780$
and $vuds\_efdcs\_530029038$) we use the sum of the components for
$DEIMOS\_COSMOS\_881725$, while we use the fluxes of the central
targets for $vuds\_cosmos\_5101209780$
and $vuds\_efdcs\_530029038$ since
the companions are likely separated objects (see Appendix D.2
and Table D.1 in \citealt{2020A&A...643A...2B} and \citealt{2020A&A...643A...7G}).

We anticipate here that in Sec. \ref{dmd_uv_sec} we will perform
a stacking analysis on the rest-frame 157 {$\mu$}m ALMA continuum images, including both the 23 individual continuum detected
sources and the non-detections. This will be essential to recover the
average properties of the ALPINE population.

 In the ALPINE survey, 57 sources were also serendipitously detected at
860-1000 $\mu$m (ALMA band-7) and we refer to \cite{2020A&A...643A...2B}
for the description of the detection procedure and the delivery of the non-targets
catalogue. In this case, given the absence of a prior, a conservative threshold of S/N>5 has been applied.
We refer to \cite{2020A&A...643A...8G} for the detailed
characterisation of the sources, done by searching for
counterparts of the serendipitous targets in all the available multi-band and
photo-z catalogues, and also performing new photometry. The resulting blind
sources are distributed in the wide photometric redshift range
(0.5-6) computed including UV, optical and IR  data (for only 4 sources
it was not possible to assign a redshift, see Sect. 3.2.2 in
\citealt{2020A&A...643A...8G}). For the aims of the present work (see
Sec. \ref{dmd_fir_sec}) we will consider the subsample of 10 galaxies in the redshift range
corresponding to the ALPINE targets ($4.1<z<5.9$). All the 10 sources
are detected in the COSMOS field.

\section{Dust mass estimates}
\label{dust_sec}

\begin{figure}
\includegraphics[width=0.5\textwidth,angle=0]{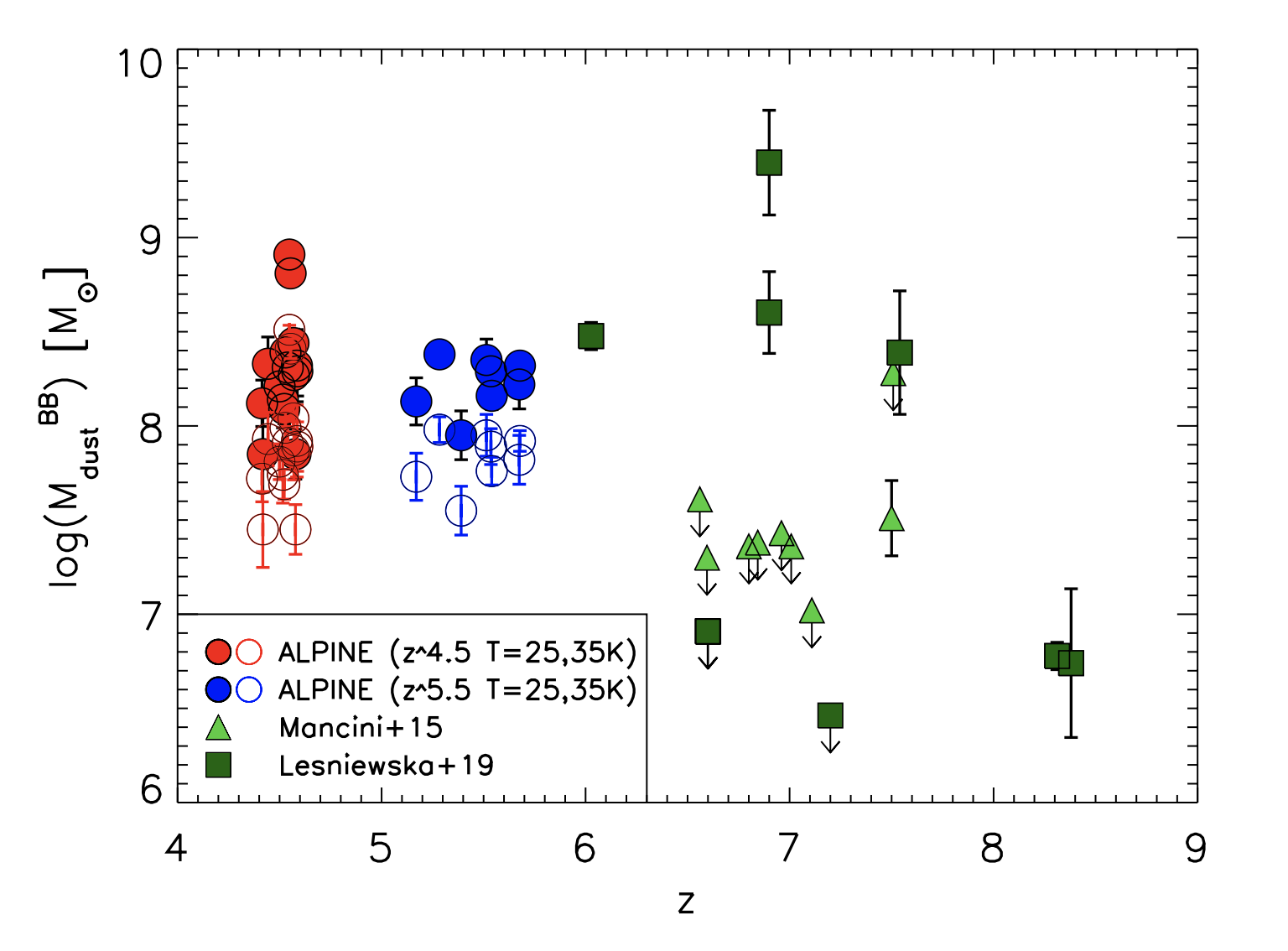}
\caption{ Dust masses of the ALPINE continuum-detected galaxies (filled
    circles) plotted as a function of $z$ and computed assuming a MBB spectrum with 
    T=25 K and $\beta=1.8$ (see
    Sec. \ref{dust_sec}). The empty circles correspond to T=35 K. Blue
   and red points show detections at 4$<z<$5
  and 5$<z<$6, respectively. For comparison, we show other 
  compilations of dust masses at high-$z$ (green triangles: \citealt{2015MNRAS.451L..70M}; dark-green squares: \citealt{2019A&A...624L..13L}).}
\label{mdust_comparison_fig}
\end{figure}

We derive the dust masses using a single Modified Black Body (MBB) curve,
under the approximation of an optically thin regime
(see \citealt{2013A&A...552A..89B}),

\begin{equation}
\label{eq_mdust}
M_{dust}=\frac{D_{L}^{2}S_{\nu_{obs}}}{ (1+z)k_{\nu}B_{\nu}(T)}
\end{equation}

\noindent where ${\nu}$ and ${\nu_{obs}}$ are the rest-frame and
observed frequencies ($\nu={\nu}_{obs}(1+z)$), $B_{\nu}(T)$ is the Planck function, $D_{L}$ the luminosity distance, $k_{\nu}$ the grain absorption cross section per unit mass and $S_{\nu_{obs}}$ is the observed flux corresponding to a rest-frame frequency at which the dust can be considered optically thin.

We assume a power-law for $k_{\nu}$,
$k_{\nu}=k_{\nu_{0}}(\frac{\nu}{\nu_{0}})^{\beta}$ cm$^2$g$^{-1}$. We adopt $k_{\nu_{0}}=4$
cm$^2$g$^{-1}$ with $\nu_{0}$=1.2 THz (see \citealt{2013A&A...552A..89B})
and $\beta=1.8$,  i.e. the Galactic value from the Planck data
(\citealt{2011A&A...536A..25P}).

  For the temperature of the dust, we assume T= 25 K.
Considering the rest-frame frequency assumed to compute $S_{\nu_{obs}}$, we have considered 1.2 THz, corresponding to 
$\lambda_{rest}=250$ $\mu$m (see \citealt{2014A&A...562A..67G}).  At this wavelength our galaxies are expected to be optically thin based on the recent results
obtained by \cite{2020MNRAS.498.4192F} from the SEDs of 4 main-sequence $z{\sim}5.5$ galaxies (2 contained in the ALPINE dataset), observed with ALMA in 3 different bands and for which constraints 
on the Rayleigh-Jeans part of their spectra were obtained. Our assumption is valid considering also the wavelengths
  at which  more dusty/extreme sources than the ALPINE ones become
  optically thick as SMGs ($\lambda_{rest}{\lsimeq}200$ $\mu$m, \citealt{2011ApJ...732L..35C}, \citealt{2013Natur.496..329R}; $\lambda_{rest}{\lsimeq}100$ $\mu$m, \citealt{2017ApJ...839...58S}).

Below we discuss our choice of a single cold component. Morever, we discuss the
uncertainties on the derived dust masses related to three (non-independent) factors: the `band-conversion', linked to the extrapolation of the 250 $\mu$m flux
from higher frequencies (i.e. 157 $\mu$m), the emissivity $\beta$ and the temperature T.

In galaxies, dust is located primarily in
two components, a warm one (20$<$T$<$60 K) associated to the Photo-Dissociation
Regions (PDRs) and a cold one associated to the diffuse ISM medium
(T$<30$ K, see \citealt{2007ApJ...657..810D}).
 The adopted method relies on the assumption that the diffuse cold dust component accounts for the bulk of the dust budget. This has
 been shown to be true in the local Universe (see \citealt{2017A&A...602A..68O}), where, given the high-quality
photometric data, a sophisticated fit has been performed showing how the contribution of the warm
component accounts for a low fraction of the total dust budget
($\sim$1 \% and reaching the highest value of 4$\%$ for starburst
galaxies). At higher redshift, up to $z{\sim}$2, the validity of this approximation has
been advocated by \cite{2014ApJ...783...84S} (see also
\citealt{2016ApJ...820...83S}, \citealt{2017ApJ...837..150S}) in support to the correlation between the gas masses derived
from the sub-millimeter fluxes (directly linked to the dust masses under the
optically thin approximation) and the gas masses derived from CO
observations, under the assumption of a dust-to-gas ratio (see also the
recent work of \citealt{2019ApJ...880...15K}).

At the redshift of the ALPINE sources, $z{\sim}$5, there is no
systematic study that compares the molecular gas masses derived from the sub-millimeter fluxes and the CO
measurements, because the CO line is hard to get even with the
most up-to-date interferometers (so far in normal main sequence
galaxies there are only two detections, \citealt{2018ApJ...863L..29D},
\citealt{2019ApJ...882..168P}). Moreover, the CO-based ALMA gas mass estimates rely on strong
assumptions about the CO excitation mechanism to extrapolate from the high-J CO levels accessible with
ALMA the CO(1-0)
fundamental line. To overcome this issue, in
\cite{2020A&A...643A...5D} the [CII] line has been used as tracer of
the molecular gas masses for the ALPINE galaxies. By using the
relation from \cite{2018MNRAS.481.1976Z}, the molecular gas masses have been derived from the [CII] luminosities and compared with masses
inferred from dynamical masses and sub-mm fluxes under the approximation of a unique
dust component at T= 25 K and $\beta$=1.8. The good correlation shown by the three tracers
(see Sect. 3 and Figs. 3 and 4 in \citealt{2020A&A...643A...5D}) supports our assumptions. A cold dust component, dominating the dust budget, has been recently
considered up to very high-$z$ also by \cite{2020ApJ...892...66M},
taking advantage of the ALMA LP survey at 1.2mm (ASPECS,
\citealt{2016ApJ...833...67W}, \citealt{2019ApJ...882..138D},
\citealt{2020ApJ...897...91G}), where the cold component
approximation has been applied up to $z{\sim}$4.5 for detected sources and
up to $z{\sim}$5.5 for stacked data (see also the recent review from \cite{2020arXiv200400934H} and
references within).

We underline that our assumed dust temperatures are `mass-weighted' mean
temperatures which should not be confused with the `luminosity-weighted'
mean temperatures, more related to the global SED shape (see
\citealt{2019MNRAS.489.1397L} for a detailed discussion on the
different dust temperature
definitions). This means that our temperatures are not comparable with
the warmer temperatures obtained by \cite{2020A&A...643A...2B} by
fitting the  $\lambda_{rest}>40 \mu$m far-IR ALPINE
stacked SED (T$=43\pm5~K$, see Sect. 4 in
\citealt{2020A&A...643A...2B} paper) or with the dust peak temperature range (30-43 K) as found by \cite{2020MNRAS.498.4192F}.
 Similar considerations are valid also when comparing the cold
 `mass-weighted'  temperature with predictions from
theoretical models. In \cite{2020MNRAS.497..956S} (but see also
\citealt{2018MNRAS.474.1718N}, \citealt{2019MNRAS.489.1397L}) the authors show with an
analytical and physically motivated  model that the dust 
located in high-$z$ (>5) Giant Molecular Clouds (GMCs) is warmer (T$\sim$60~K) than locally,
since theyh are characterized by a more compact structure than
their local analogues. The warm dust located near star-forming 
regions has a strong radiative
efficiency, therefore it determines the shape of the SED but it does not represent the bulk of the dust mass, expected to be at
lower temperature (T$\sim$20-30 K, \citealt{2020MNRAS.497..956S}).

 Considering the uncertainties related to the band conversion, we rely
on the simulations performed by \cite{2018ApJ...867..102P}. In
\cite{2018ApJ...867..102P},  the 850 $\mu$m luminosities of 
simulated SEDs are compared with the ones that would have been estimated assuming a MBB approximation with
$\beta$=1.8 and T=25K (as in the present analysis), under different
observational set-ups. In particular, the scenario of observing only with ALMA Band 7, as in our case, has been considered (see Fig. 5
in  \citealt{2018ApJ...867..102P}). The inferred luminosities
are always greater than the "real" ones, with 
differences of the order of $\sim$20\% at z${\sim}$5 
(and up to 30\% for the most extreme
halos). These mis-matches are due to the choice of the parameters
($\beta$ and T) used to infer the 850 $\mu$m flux from data at higher
frequencies and from the fact that real SEDs are not single-temperature blackbodies.
We stress that the results of \cite{2018ApJ...867..102P} are obtained at 850
$\mu$m (and not at 250 $\mu$m), but are valid  also in the present  case.
In fact, we have tested that 
our dust masses do not change if one considers 250 or 850 $\mu$m and if the
flux $S_{\nu_{obs}}$  in Eq. 1 is derived 
by means of an extrapolation at higher frequencies. 

Considering the emissivity parameter $\beta$ and the temperature T, we
test how the dust masses would change by changing individually these
parameters, by freezing the others. For the emissivity $\beta$, as
stated before, we assume $\beta$=1.8, i.e. the Galactic value from the Planck data
(\citealt{2011A&A...536A..25P}), again in agreement with
\cite{2020MNRAS.498.4192F}. Theoretical models predict a wider range for $\beta$
(i.e. \citealt{2011piim.book.....D}) and therefore we test how extreme values of $\beta$=1.5 ($\beta$=2)
would affect our dust mass estimates finding an increase (decrease) of the
dust masses by a negligible factor of less than ${\sim}15\%$ for $\beta$=1.5
($\beta$=2) (see also
\citealt{2020ApJ...892...66M}). For the temperarure of the cold dust, we explore a range of values, from T=20 K up to T=35 K. We do not consider temperatures lower than
  T=20 K, given the floor imposed by the cosmic microwave background (at {z=5-6}, T$_{CMB}$=16-19 K, respectively, see \citealt{2013ApJ...766...13D}).  A temperature of T=20 K  would increase the dust masses by a factor within the uncertainties considered (20\%), while assuming  T=35 K 
would produce a decrease of the order of 60\%.  
The dust masses derived at T=25 K are our fiducial ones, but we will discuss also the consequences of a 10 K warmer dust. \\

 In Table \ref{table1} the dust masses for the 23 continuum-detected
sources are reported. We calculate the error bars on the dust masses by summing in quadrature the relative
  uncertainties on the 157~$\mu$m continuum flux (${\sim}25\%$),
  the systematic uncertainties due to the $\beta$ emissivity  (${\sim}15\%$) and to the band conversion 
  emissivity (${\sim}30\%$). The combined errors give a
  typical uncertainty of 0.2-0.3 dex, expressed in logM$_{dust}$.

 The dust masses of the galaxies of the ALPINE sample are shown as a function of redshift in Fig. \ref{mdust_comparison_fig}
  (red and blue filled circles).  
For most of our galaxies we find dust masses in the range log(M$_{dust}/M_{\odot}$)=7.8-8.4, with the exception of two systems showing 
significantly higher values (log(Mdust=M)=8.8 and log(Mdust=M)=8.9 for $DEIMOS\_COSMOS\_818760$ and $DEIMOS\_COSMOS\_873756$, respectively). 
These two sources are the only targets presenting an ALMA continuum flux $>1$ mJy (see Table B.1. in  \citealt{2020A&A...643A...2B}). The 
empty circles in Fig. \ref{mdust_comparison_fig} correspond to the
estimates obtained assuming for the dust a warmer temperature of T=35
K. In Fig. \ref{mdust_comparison_fig} we also report two collections of high-z ($z{\sim}6-7$)  dust mass measurements compiled in \cite{2015MNRAS.451L..70M} and \cite{2019A&A...624L..13L}. 
Altogether, the values from these works, which include both detections and upper limits, show a $\sim$ 3 dex scatter, hence much larger than
the one obtained for our galaxies.

\begin{figure*}
\center
  \includegraphics[width=0.8\textwidth,angle=0]{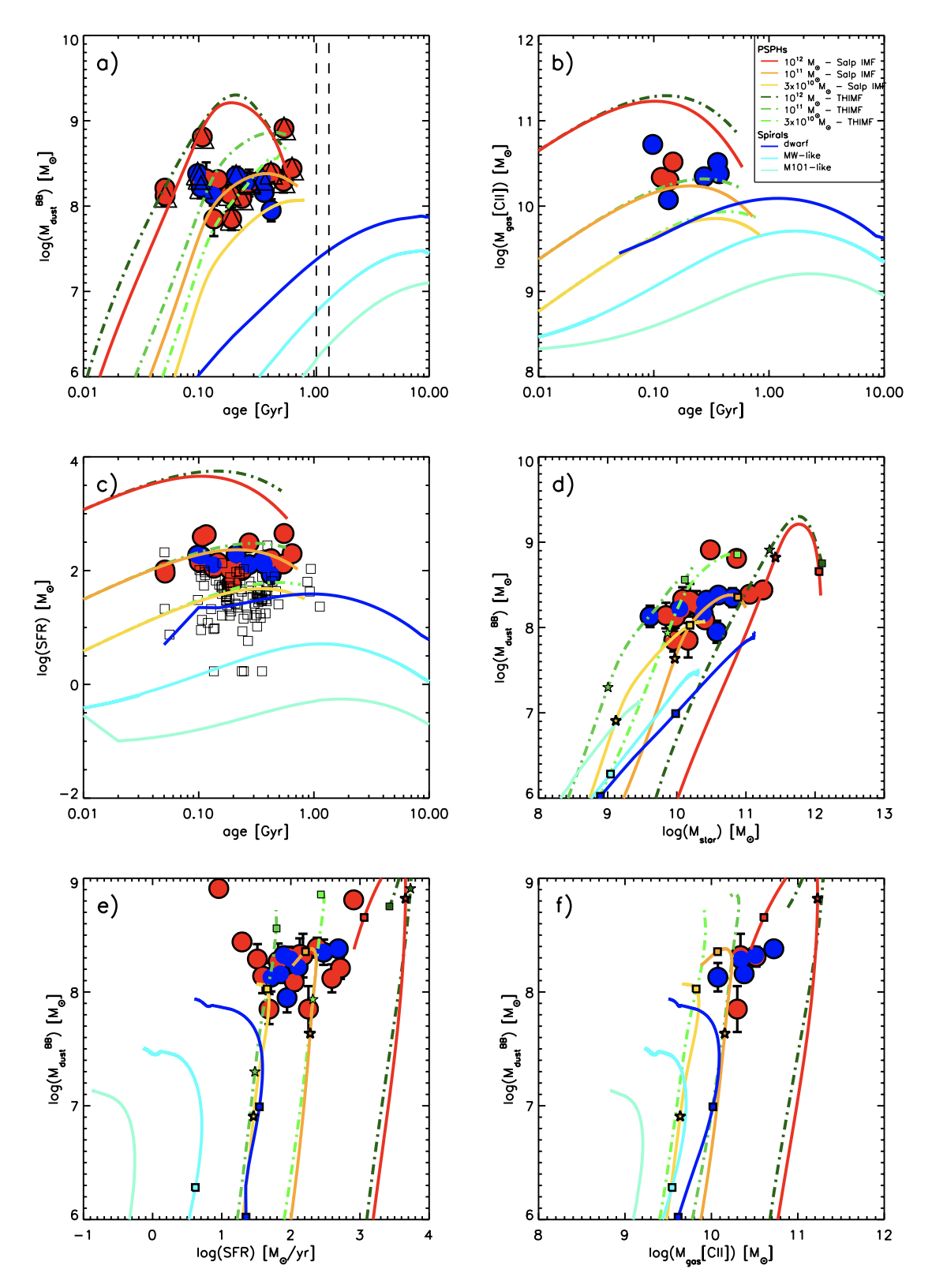}
  \caption{Top panels: M$_{dust}$ versus age (panel a) and 
    M$_{gas}$ versus age (b).
    Middle panels: SFR versus age (c) and M$_{dust}$ versus M$_{\star}$ (d). 
    Bottom panels: M$_{dust}$ versus  SFR (e) and M$_{dust}$ versus M$_{gas}$ (f). 
    M$_{dust}$ has
   been computed by means of a MBB fit and assuming T=$25~K$ and $\beta=1.8$. Blue
   and red points show detections at 4 $<z<$ 5
  and 5 $<z<$ 6, respectively. In panel a) the triangles around the points indicate galaxies classified as mergers or dispersion-dominated
  objects (\citealt{2020A&A...643A...1L}), whereas the black vertical dashed lines indicate the age of the Universe
at redshift 4.5 and z=5.5 (corresponding to an age of the Universe of 1.35 Gyr and 1.05 Gyr, respectively).
  The empty squares in panel c) represent the ALPINE galaxies
  non-detected in continuum. The curves represent the chemical evolution models described in Sect. \ref{chemi_sec}. The yellow, orange and red
  solid lines represent proto-spheroid models of baryonic mass
  3${\times}10^{10}$, 10$^{11}$ and 10$^{12}$ M$_{\odot}$,
  respectively, computed with a standard Salpeter IMF (\citealt{1955VA......1..283S}).
  The light green, green and dark green dot-dashed lines are for three PSPHs with the same baryonic masses
  as above but characterised by a top-heavy IMF
  (\citealt{1998MNRAS.301..569L}). The  light cyan, cyan and blue solid lines represent models for a  dwarf spiral, an
  intermediate-mass spiral and an M101-like spiral, respectively
  (\citealt{2017MNRAS.465...54C}). 
The stars and squares plotted along each curve and with the same colour mark the evolutionary times of 0.1 and 0.5 Gyr, respectively.}
\label{scaling_fig}
\end{figure*}

\section{Scaling relation of ALPINE continuum detected sources  }
\label{physics_sec}

In this section we combine the dust masses estimates
with other physical information obtained
for the ALPINE targets, and we compare the
observed dust scaling relations with
predictions from chemical evolution models in order to understand
the nature and to probe the evolution of MS-selected galaxies at z$\sim$4-5.
In the following section we present first a brief description of the
chemical models and then the comparison of the observed dust scaling relations with the model predictions.

\subsection{Chemical evolution models}
\label{chemi_sec}

The chemical evolution models considered here 
describe galaxies with different star formation histories, namely
proto-spheroidals (PSPH) and spirals.
The models including evolution of interstellar dust grains were presented in \cite{2008A&A...479..669C} 
 (see also \citealt{2009MNRAS.394.2001S}, 
\citealt{2011A&A...525A..61P}, \citealt{2014MNRAS.438.2765C}).
Chemical evolution models calculate the evolution of the abundances of
various chemical elements and dust in the interstellar medium, generally starting from simple prescriptions
to describe basic processes regulating galaxy evolution. 
In the past, these models have been extensively tested and
used to interpret observational, dust-related  scaling relations in
local and distant galaxies (see also \citealt{2017MNRAS.465...54C}).

In the models used here, PSPH represent the precursors of local elliptical galaxies. 
PSPH form from the rapid collapse of
a gas cloud of primordial composition, which triggers an intense starburst.
The collapse is described by an exponential infall law, with a characteristic e-folding
time which depends on the mass of the galaxy (\citealt{2014MNRAS.438.2765C}) and with typical values in the range 
$\sim 0.2-0.5$ Gyr (\citealt{2020MNRAS.494.2355P}).  
The starburst continues until the thermal energy of the interstellar medium, computed as 
the cumulative energy deposited in the ISM by type II and type Ia SNe, equals the binding energy
of the gas, computed taking into account the gravitational 
contribution of the gas, of the stars and of the dark matter halo, 
assumed ten times more massive than the initial gas mass (\citealt{1994A&A...288...57M}).
When this occurs, star formation is immediately interrupted, and the galaxy is assumed to instantaneously
eject all the residual gas. 
The models describe three PSPH of final baryonic mass $3\times
10^{10}~M_{\odot}$, $10^{11}~M_{\odot}$ and $10^{12}~M_{\odot}$.

In this work, we also consider a set of chemical evolution models for spiral galaxies. 
Each model consists of several independent rings, 2 kpc wide, each of them representing a separate region of a disc.
Also in spiral galaxies the formation of the disc is described by an exponential infall, but with much longer, radius-dependent timescales, 
ranging from $\sim1$ Gyr in the innermost parts up to $\sim10$ Gyr in the outskirts, according to the 'inside-out' scenario
(\citealt{1989MNRAS.239..885M}). 
The final stellar masses of the spirals range from $2\times 10^{9}~M_{\odot}$ to $10^{11}~M_{\odot}$.

In all our models, star formation is modelled by means of a
\cite{1959ApJ...129..243S} law,  
with a higher efficiency in more massive objects, a  behaviour known
as galactic downsizing (e.g., \citealt{1996AJ....112..839C}; \citealt{1994A&A...288...57M}, \citealt{2020A&A...642A.113S}). 

The stellar initial mass function (IMF) assumed for spirals is the one of \cite{1986FCPh...11....1S}, 
a two-slope power-law characterised by a steeper slope ($\alpha=-1.7$) with respect to the Salpeter ($\alpha=-1.35$) at stellar mass values $>2  M_{\odot}$.

For the IMF of PSPH we test two different 
forms, i.e. the \cite{1955VA......1..283S} IMF and the one of
\cite{1998MNRAS.301..569L}, a top-heavy IMF (THIMF).
The Larson IMF has the form $\phi (m) \propto m^{-1.35} exp(-m_c/m)$, 
and assuming a characteristic mass value $m_c=1.2 M_{\odot}$, it is top-heavy, i.e. richer in massive stars, with respect to the Salpeter IMF
(see Calura et al. 2014).
In all cases, the IMF is assumed to be constant in time.

The use of a steeper IMF in discs 
is motivated by the results by \cite{2005ApJ...625..754W},  
who have shown that, since the disc stellar population is mostly made
by dissolving open stellar clusters, the disc IMF must be significantly steeper than the 'canonical' cluster IMF (with $\phi(m) \propto m^{-1.35}$ for stellar masses above $\sim~1~M_{\odot}$) 
as the former results from a folding of the latter with the star cluster mass function, which in general is a single slope power law 
(e. g., \citealt{2003ARA&A..41...57L}). 

The choice of a different IMF in different models 
is also motivated by the requirements of reproducing the local constraints, including the abundance pattern observed
in the Milky Way disc for spirals (e. g., \citealt{2019A&A...623A..60S}) 
and the abundance ratios and metal budget in local ellipticals (e. g. \citealt{2004MNRAS.350..351C}; \citealt{2019MNRAS.483.2217D}).
For the stellar mass and SFR, the conversion between a standard IMF such as the one of \cite{2003ApJ...586L.133C} and
    the one of \citep{1955VA......1..283S} amounts to $\sim$0.2-0.25 dex. The same is true also when a \cite{1986FCPh...11....1S} is considered. 
Therefore, the adoption of different IMFs produces variations which are in general much smaller than the range shown by the ALPINE sample. 
Overall, this implies that the impact of the choice of the IMF on the
results of Fig.  \ref{scaling_fig} is marginal. 

The models include dust production in stars, occurring in core-collapse SNe and intermediate-mass stars, restoring
significant amounts of dust grains during the asymptotic giant branch (AGB) phase. 
The set of metallicity-independent dust condensation efficiencies considered here are from \cite{1998ApJ...501..643D}.
Also dust destruction in supernovae shocks and dust growth in the ISM are taken into account. 
The prescriptions for these processes are described in
\cite{2008A&A...479..669C} (see also \citealt{1998ApJ...501..643D}).

\subsection{Dust scaling relations}
\label{scaling_sec}

In Fig. \ref{scaling_fig} the dust masses are reported along with other physical quantities of the ALPINE continuum detected sources: the stellar mass (M$_{stars}$), the star-formation
rate (SFR), the age, and the gas
mass (M$_{gas}$). The stellar mass, the SFR and the age values have been computed by
\cite{2020ApJS..247...61F} adopting a \cite{2003ApJ...586L.133C}
IMF and using the LePhare SED-fitting code (\citealt{1999MNRAS.310..540A}, \citealt{2006A&A...457..841I}). We are aware that the age is
one of the most uncertain parameters of the SED-fitting analysis 
(i.e., \citealt{2017A&A...602A..35T}). However, as we will discuss later, 
the exact values found for this parameter do not influence significantly our results.

For consistency
with the models, we rescale  M$_{stars}$ and SFR to a
\cite{1955VA......1..283S} IMF by multiplying the quantities by a factor of 1.7 (see \citealt{2014ApJS..214...15S}).
Regarding M$_{gas}$, we adopt the values from \cite{2020A&A...643A...5D}, obtained
from the [CII] luminosities using the \cite{2018MNRAS.481.1976Z} relation, with a typical
uncertainty of $\sim$0.3 dex.
In Fig. \ref{scaling_fig} we compare estimates derived for the galaxies
of our sample with results obtained with the chemical evolution models presented in Sect. \ref{chemi_sec}.
The solid circles represent the inferred quantities (red and
blue for galaxies at $z{\sim}4.5$ and  $z{\sim}5.5$, respectively), while the
curves represent the PSPH and spiral models (see caption of Fig.\ref{scaling_fig} for further details). 

The evolution of the dust mass as a function of age (a) panel) shows
that the inferred M$_{dust}$ values are in good agreement with those obtained 
with PSPH models at epochs comparable to the ages of the detected
systems.
 
Typical dust mass values of $\sim10^8~M_{\odot}$ as derived in the
ALPINE sample of continuum detected sources are achievable already at early times (typically $\sim 0.1-0.5$ Gyr) in PSPH.
On the other hand, in the case of spirals only the most massive 
model reaches comparable values after several Gyr of evolution (blue line in the top-left panel),
whereas the other models show much lower M$_{dust}$ values at any epoch. As already said, this result is robust against the
exact value of the ages of the ALPINE galaxies. In fact, in the extreme and
unlikely case that the ages are all systematically underestimated, 
they can not be larger than the age of the Universe at the 
redshift of our sample (vertical dashed lines in Fig.\ref{scaling_fig}, panel a)).

This seems that the PSPH models are the only ones that can cover the region occupy by the 
observed galaxies. The other models need $>1$ Gyr to produce the
M$_{dust}$ that are observed. This finding is due to an overall much faster, stronger 
evolution shown by PSPH with respect to spirals, confirmed also by
their star formation history (panel (c) in Fig.\ref{scaling_fig}). The SFR vs age plot shows that
most of the SFRs measured in ALPINE galaxies
are consistent with the values shown by the intermediate- and low-mass PSPH models (orange and yellow lines),
with very few of them similar to the values shown by the most massive spiral models.
This plot also shows a limited time interval in which the SFR values of the low-mass PSPH are
very similar to those shown by the massive spirals at early times ($<1$ Gyr),
but in other plots these two models show a rather distinct behaviour, 
such as in the M$_{dust}-$Age and M$_{dust}-$M$_{stars}$ relations. 
A couple of galaxies with particularly high M$_{dust}$ are reproduced
by the PSPH models with a THIMF (green dashed-dotted lines). 
On this topic, a particularly high dust yield (defined as 
amount of dust per unit stellar mass, or specific 
dust mass) in star-forming galaxies at high redshift was found also
in other works (\citealt{2017MNRAS.465...54C} and references therein), 
and several other reasons were proposed, including, beside the adoption of
a THIMF (\citealt{2011A&ARv..19...43G}), also an enhanced dust accretion rate
(\citealt{2011A&A...525A..61P}; \citealt{2011MNRAS.416.1916V}).

Another explanation for this result invokes a substantial revision
of our current knowledge of both dust production and accretion mechanisms, which might be needed in the case that
neither destruction in SNe shocks nor accretion plays a significant role in regulating the dust mass budget in galaxies. 
In a recent analysis of the dust mass budget in local and distant star-forming galaxies, \citealt{2018ApJ...868...62G} showed that 
the observed dust masses can be accounted for if the majority of the dust is formed by SNe on short timescales, i.e. during the most recent star formation episodes. 
This does not exclude that other processes, such as grain growth and destruction, may be at play in regulating the dust budget but,
if their effects are non-negligible, they need to balance each other and occur on comparable timescales. If this is the case,
several parameters related to the gas microphysics and sometimes different local thermal conditions 
have to be tuned to yield comparable destruction and growth rates. 
A most direct interpretation of this results is that stardust production is dominant with respect to other processes. 
In such case, a high dust yield can be accommodated assuming a THIMF.

Also the analysis of the cold gas budget, as traced by the molecular gas content and 
indicated by the M$_{gas}$ vs age, supports the result that 
ALPINE galaxies show properties mostly similar to PSPH.

A further indication supporting our conclusion, is the galaxy
classification of the ALPINE targets, performed by
\cite{2020A&A...643A...1L}, from a preliminary visual inspection
analysis of the [CII] data cubes together with the large wealth of
ancillary data (mainly HST F814W images, see
\citealt{2007ApJS..172..196K}, \citealt{2011ApJS..197...36K}). The majority of the ALPINE [CII]-detected galaxies have a disturbed morphology (60$\%$ including  mergers or
extended and dispersion dominated systems) while only 13.3\% show a
rotating disc (the remaining 10.7\% and  16$\%$ being compact or too
faint to be classified). The fraction of galaxies with disturbed
morphology become even larger (${\sim}70\%$) when
considering only the continuum-detected sources. This does not exclude that the 
   ALPINE galaxies might be compatible with progenitors of MW-like galaxies, whose earliest assembly and star-formation activity 
   is expected to occur mostly in their spheroidal components.

Moreover, our conclusion is fairly robust against the fact that is based on
only 23 ALPINE continuum detected galaxies. In Fig. \ref{scaling_fig} (panel c)), we report
the SFR as a function of the age also for the ALPINE galaxies not detected in continuum (empty squares). These sources, as expected, are
 less extreme, characterised by a lower SFR in
 comparison to the continuum detected ones ($<$SFR$>\sim$56 M$_\odot$
 yr$^{-1}$ and $<$SFR$>\sim$180 M$_\odot$ yr$^{-1}$, respectively). Nevertheless, they still occupy a parameter
space between the evolutionary tracks of the PSPH and the most massive 
spiral model. 

In conclusion, our study of the scaling relations and our comparison with chemical evolution models indicates that ALPINE galaxies
mostly show dust masses and SFR values expected in young, star-forming  PSPH. In most cases, a Salpeter IMF
is sufficient to account for the observed dust masses, i.e. in general there is no particular need to invoke mechanisms such as a top-heavy
IMF to enhance dust production, as seen e. g. in a sample of starbursts observed by Herschel (Calura et al. 2017).
Our models for disc galaxies show a slower buildup of the dust mass and, at these epochs, 
fail to account for the observed dust masses.

Our study strongly outlines the need for statistical samples of
 galaxies at high-$z$ with the SED properly sampled in the far-IR/sub-mm
 part of the spectrum, in order to better derive the dust properties
(mainly the temperature and the mass). This will be achieved by exploiting the synergies between sub-mm/mm
 (i.e. ALMA, NOEMA) and
 far-IR facilities, 
with features similar to the SPace Infrared telescope for Cosmology
and Astrophysics (SPICA\footnote{http://spica-missions.org},
\citealt{2018PASA...35...30R}) or to the Origins
Space Telescope (OST\footnote{http://origins.ipac.caltech.edu})
which, altogether, will enable a full characterisation of the spectrum
from the Wien (FIR) up to the Rayleigh-Jeans regimes (sub-mm/mm).

This will allow us to 
  improve our understanding of early dust production and to increase
  the current samples of galaxies with measured dust masses at
  high-redshift. In the forthcoming future, the James Webb Space Telescope (JWST, \citealt{2009ASSP...10....1G}), working in
  the near-IR and mid-IR regimes, will allow us to better
constrain the stellar masses, ages, metallicities and star
formation histories of galaxies at very high-redshift.

\begin{figure}
\includegraphics[width=0.5\textwidth,angle=0]{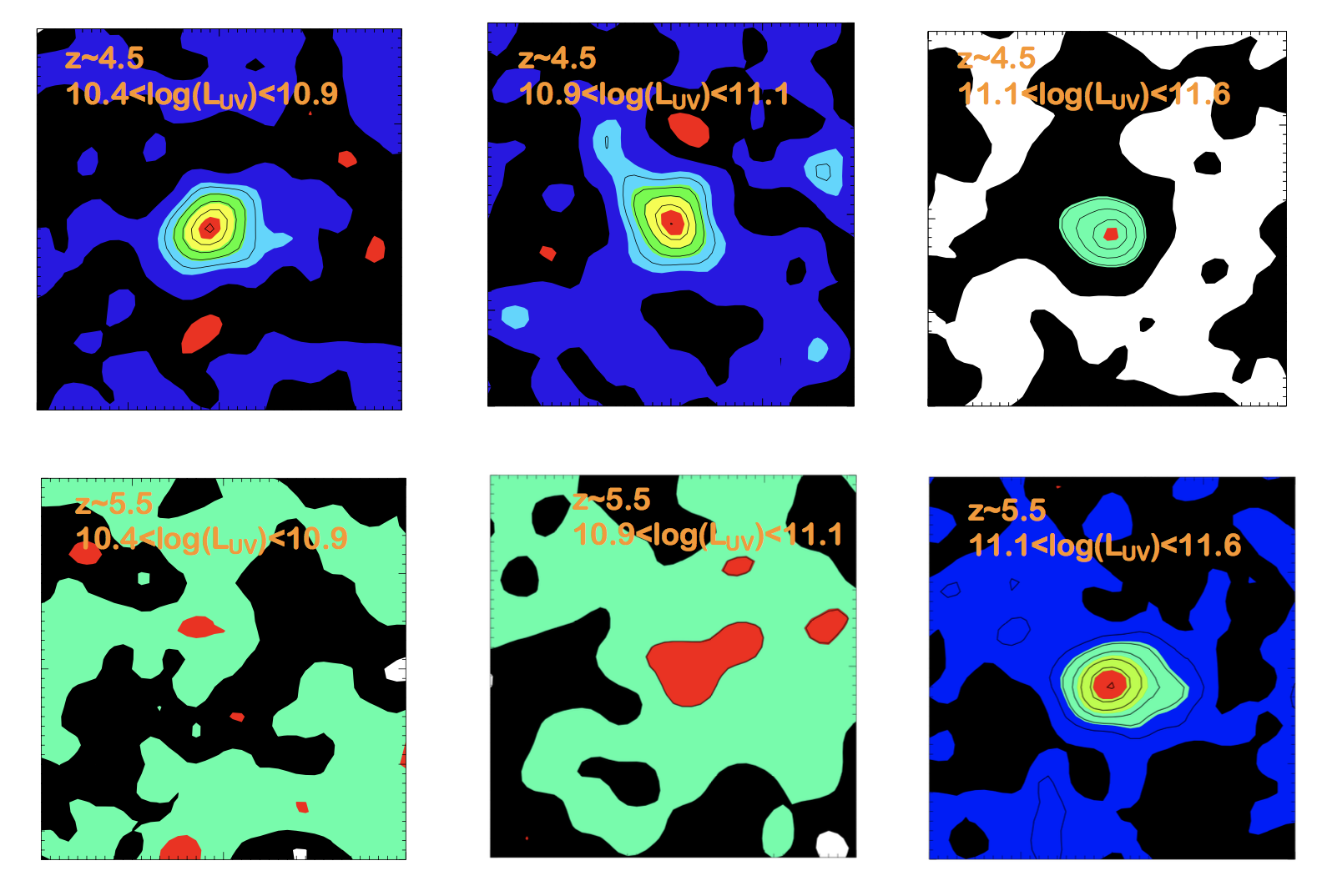}
\caption{$6^{\prime\prime}{\times}6^{\prime\prime}$ cutouts of L$_{UV}$
  binned stacks of ALMA continuum images used to derive the stacked dust
  masses. The upper and lower panels show stacks of galaxies at
  $z{\sim}$4.5 and $z{\sim}$5.5, respectively. Black solid contours show
  2,3,4,5 $\sigma$. While in the $z{\sim}$4.5 bins all stacks have a
  clear detection at $>4{\sigma}$, in the $z{\sim}$5.5 bins the stacks
  have a clear detection only at the highest L$_{UV}$ bin. In Table
  \ref{table2} the results of the stacking analysis are summarised,
  including the number of sources stacked in each bin of L$_{UV}$.}
\label{stacked_fig}
\end{figure}

\section{Dust mass density at $z{\sim}$5}
\label{dmd_sec}

In the following section, we  derive an estimate of the dust mass
density (DMD) in the redshift range of the ALPINE survey ($4.3< z < 4.6$ and $5.1< z < 5.9$).  Taking into account that
the ALPINE is a targeted survey of pre-selected galaxies, we divide the
derivation of the DMD in two steps: first, we estimate the
contribution of the UV-selected galaxies ($\rho_{dust,UV}$), which sets a lower
limit to the DMD; second, we consider the continuum serendipitously
detected sources in the redshift bin of the ALPINE targets, trying to evaluate
the total DMD ($\rho_{dust,IR}$).

\begin{figure}
\includegraphics[width=0.5\textwidth,angle=0]{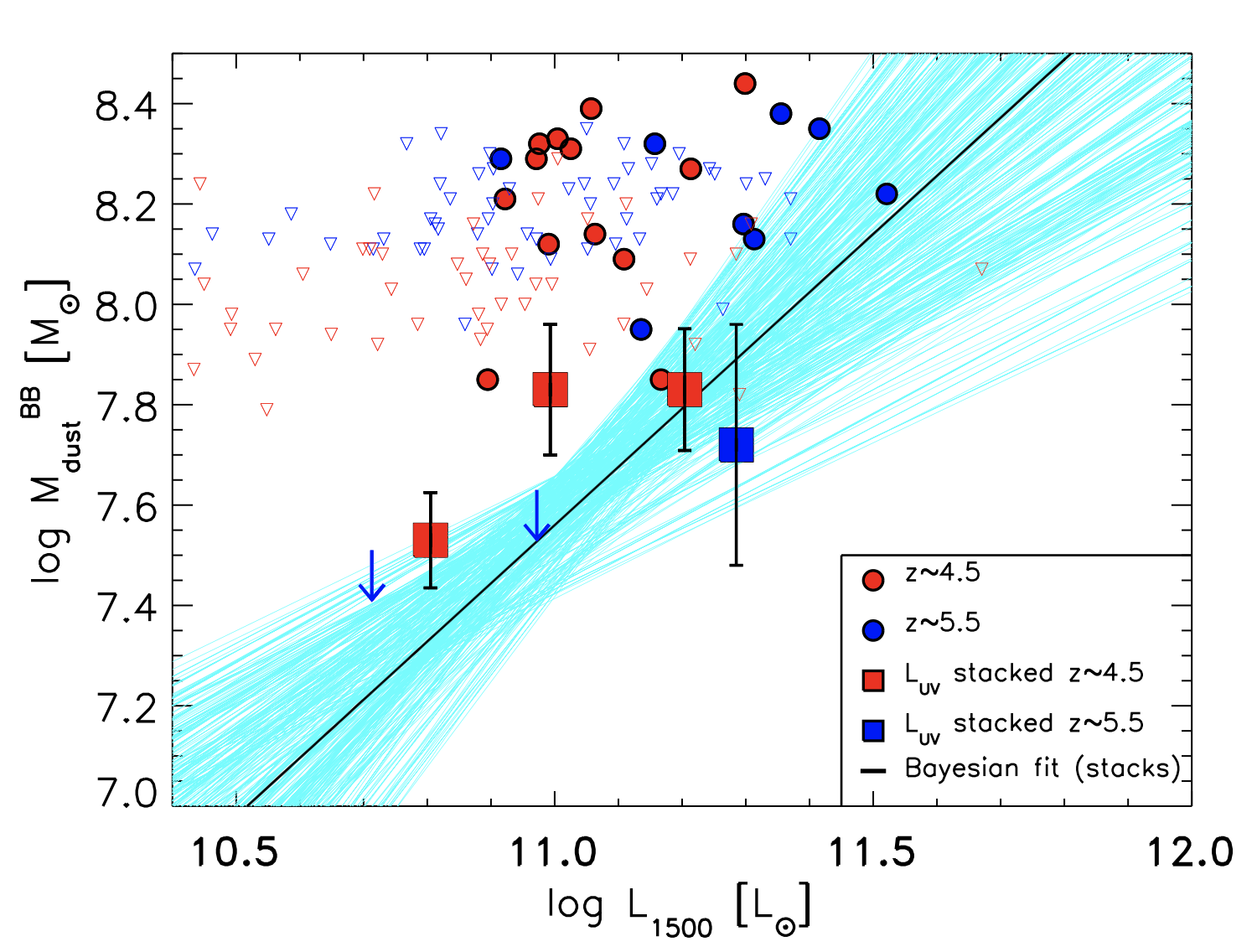}
\caption{M$_{dust}$ versus L$_{UV}$ for the ALPINE targets. Blue and red points show
  individual FIR continuum detections at $4<z<5$ and $5<z<6$,
  respectively. Open downward triangles show 3$\sigma$ upper limits of
  individual IR non-detections. Stacks with detections at $z<5$ and $z>5$, are shown as
  blue and red squares, respectively. The non-detection of stacks are indicated by a downward arrows. M$_{dust}$ from the stacked
  ALMA continuum fluxes have been estimated using the same expression
  as for the targets (Eq. \ref{eq_mdust}). The Bayesian linear
    fit of all the stacked values (detections and upper limits) is
    shown by the black line/light-blue lines, which also 
    illustrate the probability distribution of the fit at 1$\sigma$
    confidence level. The resulting best-fit is
    $log(M_{dust})=(1.16^{0.68}_{-0.57})(log(L_{UV})-11.)+(-0.11^{0.10}_{-0.15})+7.76$
    (dispersion ${\sigma}{\lsimeq}0.2$ dex, Pearson
    rank coefficient of 0.76 and
    p-value 0.08, with the Pearson coefficient computed considering all data as detections)}.
\label{mdust_versus_luv_fig}
\end{figure}

\subsection{Contribution to the DMD from UV-selected galaxies}
\label{dmd_uv_sec}

The ALPINE targets are UV-selected sources, with UV-magnitude L$_{UV}>0.5L^{\star}$ (see
Sect. \ref{data_sec}). Among the 118
ALPINE targets, 23 sources ($\sim20\%$ of the sample) are far-IR continuum
detected and, as a consequence, for deriving the average
properties of the parent UV-selected population, we need to take into
account also the contribution of the non-detected galaxies. To
this aim, we performed a stacking analysis on the ALMA maps.

We perform stacking using the $\lambda_{158{\mu}m}$ rest-frame continuum images
centered on the UV counterpart positions, including both individual
detections and non-detection in order to avoid biases. The procedure used has
been extensively described in the ALPINE paper of \cite{2020arXiv200708384K}
(see also \citealt{2020arXiv200410760F}).  Here we briefly summarise the main
points. We use median stacking on the UV counterpart
positions of both detected and non detected objects. As for the detections, we adopt a
3.5$\sigma$ threshold. In case of non detection, we use the
conservative 3$\sigma$ upper limits by adding three times the
background RMS to the local maximum (see \citealt{2020A&A...643A...2B}.)
We performed the stacking in two redshift bins, at $z<5$ and z$>5$,
and we further split the sample in bins of UV luminosity. The bins
have been chosen to have an almost equal number of sources in each
interval ($\sim20$). In Table \ref{table2} the results of our
stacking analysis are summarised, while in Fig. \ref{stacked_fig} the stacked images are
shown. From the stacks at $z<5$, we detected
significant continuum emission from all the L$_{UV}$ bins
(Fig. \ref{stacked_fig}, upper panels), while from
the stacks at $z>5$, we detected continuum emission only in the
highest L$_{UV}$ bin (Fig. \ref{stacked_fig}, lower panels).

In Fig. \ref{mdust_versus_luv_fig} the dust masses as a function of
their UV-luminosity have been reported for the continuum detected
sources (23 objects), for the
non-detections and from the stacked images. For the non-detections and
for the stacked images, the same procedure adopted for the detected
sources has been used to estimate the dust masses, considering the 3$\sigma$ flux upper limits and
the stacked results as rest-frame 158~$\mu$m continuum fluxes,
respectively. The black line represents the
linear log(M$_{dust}$)-log(L$_{UV}$) relation obtained using the
stacked values. 
We performed the fit using the method described
  in \cite{2019A&C....2900331F}, for evaluating the likelihood in
  presence of censored data. This method is a generalization of the
  statistical approach of \cite{2007ApJ...665.1489K} and is implemented in the {\it LeoPy} python
  package. The evaluation of the likelihood has been combined
  with a Bayesian tool to realize Monte Carlo Markov Chain for parameter space exploration, implemented
  in the {\it emcee} python package (\citealt{2013PASP..125..306F}).
The fit results with a poor significance level with the
coefficient constrained at 1-2$\sigma$ (Pearson rank
coefficient of 0.76 and  p-value of 0.08, estimated considering all
data as detections). Regardless of the poor significance of the
  fit, we are confident about the L$_{UV}$-M$_{dust}$ correlation, as a
  secondary effect of the relation between SFR and M$_{dust}$, in
  galaxies where  L$_{UV}$  captures most of the SFR (as expected for the ALPINE
  targets, characterized generally by low extinction values, $<(E-V)>{\sim}0.08$, see \citealt{2020ApJS..247...61F}).

  The obtained relation allows us to derive the DMD for the UV-selected
  sources. In particular, since the ALPINE sample is not a
  volume-limited sample, to derive the comoving
density of our physical property (M$_{dust}$), we use L$_{UV}$ as proxy of
M$_{dust}$, following the same approach described in a companion ALPINE paper (\citealt{2020arXiv200708384K})
where the authors derived the Star Formation Rate Density (SFRD) starting from the continuum
detected ALPINE UV-selected sources. 

 Following \cite{2020arXiv200708384K} (see also
 \citealt{2004A&A...424...23F}, \citealt{2020MNRAS.491.5073P}), we estimate the comoving
density of our physical property by convolving the volume density
of the proxy with the mean M$_{dust}$-L$_{UV}$ relation as:

\begin{equation}
 \rho_{dust,UV}=\int{ <M_{dust}>(x) \phi(x) dx}
\end{equation}

\noindent where $x$ is the L$_{UV}$ and $\phi(x)$ is the UV-luminosity
function (UVLF). In this operation we have converted the mean relation
found in log space
(see Fig. \ref{mdust_versus_luv_fig}) to the mean relation in linear space as:

\begin{equation}
<M_{dust}/L_{UV}>=10^{(<\log{\frac{M_{dust}}{L_{UV}}>+\frac{2.3\times{\sigma}^2}{2}})}
\label{eq2}
\end{equation}

\noindent where $\sigma$ is the dispersion of the data in log space as
obtained by the Bayesian fit (${\sigma}{\sim}0.2$ dex, see Fig. \ref{mdust_versus_luv_fig}).

Two main caveats could affect our method and should be mentioned: first, we
consider the ALPINE targets as a fair representative sample of the
UV-sources; second, we extrapolate to a broader L$_{UV}$ range, the log(M$_{dust}$)-log(L$_{UV}$) relation found
for $10.4<$log$(L_{UV})<11.5$. Considering the first issue, we rely 
on \cite{2020ApJS..247...61F} (their Fig. 17), where it has been shown how the
ALPINE sources occupy the same region of the parent sample in COSMOS
(rest-frame UV/optical selected at $z{\sim}$5) in the
M$_{star}-$SFR plane, leading the authors to conclude that the ALPINE sources are a fair representation
of UV-selected star-forming $z>4$ galaxies. This is not to be intended that
the ALPINE galaxies are a representation of all the star-forming
galaxies since
we are aware that the
UV-selection does not take into account ultra-dusty
galaxies. We discuss this point in Sec. \ref{dmd_fir_sec}. 
Considering the second issue, we
estimate two quantities:  the dust mass density derived by integrating the UV-luminosity only over the ALPINE UV-luminosity range 
and the dust mass density obtained by extrapolating the
relation down to fainter UV luminosity not sampled by the ALPINE
targets, and integrating over the commonly used
$0.03L^{\star}-100L^{\star}$ range ($\rho_{dust,UV}$).

Given the procedure described above, we have estimated the uncertainty on
$\rho_{dust,UV}$ combining the uncertainties affecting the 
UVLF determination and the log(M$_{dust}$)-log(L$_{UV}$)
relation. At $z{\sim}$5, the UVLF is quite well constrained, the
main uncertainty being the faint-end slope $\alpha$ (i.e. \citealt{2015ApJ...803...34B},
\citealt{2018PASJ...70S..10O}, \citealt{2020A&A...634A..97K}). We use
the derivation from \cite{2018PASJ...70S..10O} and we estimate the
associated uncertainty by varying each of the UVLF parameters ($\alpha$, L$_{\star}$ and $\Phi_{\star}$) 
within 1-${\sigma}$ (see Table 7 in \citealt{2018PASJ...70S..10O}). We find an uncertainty
of $\sim$0.2 dex in $\rho_{dust,UV}$, of the same order of the uncertainties
derived from the 1$\sigma$ dispersion of the
log(M$_{dust}$)-log(L$_{UV}$) relation. 

 The results are shown in Fig. \ref{dmd_fig} and reported in Table
\ref{table3}.

\begin{figure}
\includegraphics[width=0.5\textwidth,angle=0]{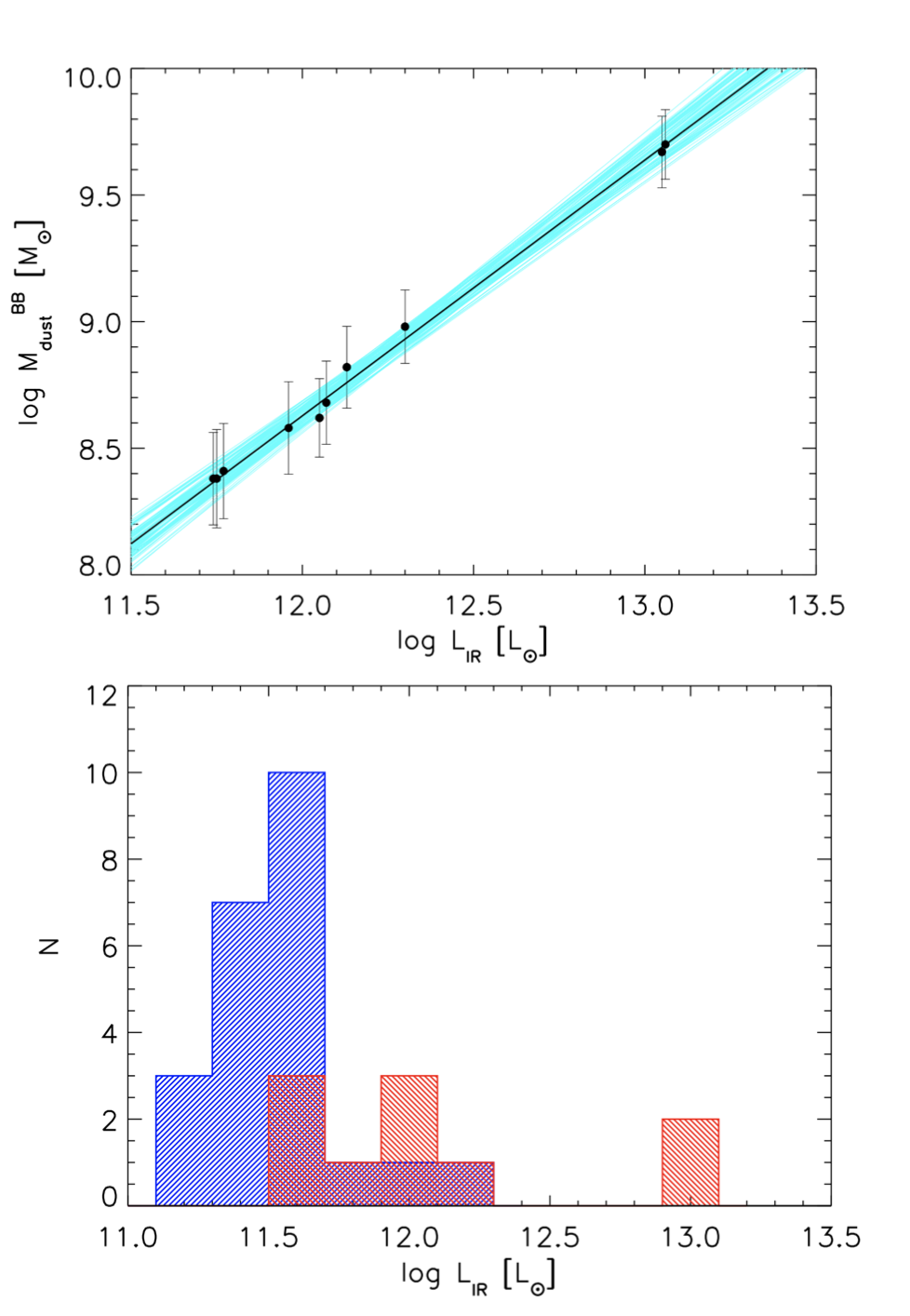}
\caption{Top: M$_{dust}$ versus L$_{IR}$ for the non-target serendipitously detected sources in the ALPINE redshift range
  (4.1$<z<5.9)$. The corresponding M$_{dust}$ have been calculated using the same expression
  as for the ALPINE targets (Eq. \ref{eq_mdust}). The Bayesian linear
    fit is shown by the black line/light-blue lines, which also 
    illustrate the probability distribution of the fit at 1$\sigma$
    confidence level. The
    resulting best-fit is
    $log(M_{dust})=(1.01^{0.11}_{-0.11})(log(L_{IR})-12.2)+(-0.0^{0.05}_{-0.05})+8.8$
    (dispersion ${\sigma}{\lsimeq}0.05$ dex, Pearson
    rank coefficient of 0.97 and p-value $10^{-5}$)}.
  Bottom: Distribution of the L$_{IR}$ for the target continuum
  detected (blue histogram, $<$log(L$_{IR})>=11.6^{+0.1}_{-0.2}$) and for the serendipitously
  detected (red histogram, $<$log(L$_{IR})>=12.1^{+0.9}_{-0.3}$ ) ALPINE sources. L$_{IR}$ have been
  estimated using the stacked SED from ALPINE analogues in the COSMOS field from \cite{2020A&A...643A...2B}. 
\label{mdust_versus_lir_fig}
\end{figure}

\subsection{Contribution to the DMD from FIR blind detected galaxies}
\label{dmd_fir_sec}

In the previous section, by integrating the UVLF over the wide
$0.03L^{\star}-100L^{\star}$ range, we have estimated the value of the
DMD of the population of rest-frame UV-selected galaxies. In the
present section, we estimate the DMD from the population of rest-frame
FIR selected galaxies. To this aim, we consider the blind continuum non-target
detections in the ALPINE survey in the redshift range
corresponding to the ALPINE targets ($4.1<z<5.9$) (see Sec
\ref{data_sec}).
For these sources we estimated the dust
masses as done for the ALPINE targets. We consider now L$_{IR}$ as proxy of M$_{dust}$ and as luminosity function of the proxy the recent determination from \cite{2020A&A...643A...8G}. Of
  the 10 sources blindy detected, 5 sources were found to have a
  spectroscopic redshift (from the detection of the [CII] emission
  line) very close to the ALPINE central targets (see
  Sect. 3.2 in \citealt{2020A&A...643A...8G}). Since these sources are
  possibly related to an overdensity (see also
  \citealt{2018A&A...615A..77L}, \citealt{2020arXiv200604837L} on this
  issue), we consider the conservative IRLF determination
  from  \cite{2020A&A...643A...8G}, where these sources have been removed.
For L$_{IR}$ we consider the values measured with the SED obtained from
stacking the photometric data-points of the ALPINE analogs galaxies in the COSMOS field from $\lambda_{rest}>40 \mu$m. In
Fig. \ref{mdust_versus_lir_fig} (Top) the values of M$_{dust}$ as a function of
L$_{IR}$ are reported together with the best-fitting relation. The fit
results with a high-level of significance (Pearson rank coefficient
0.97 and p-value of $10^{-5}$).

As for the DMD obtained for the UV-galaxy population, we estimate the comoving
density of our physical parameter ($\rho_{dust,IR}$, where now the
subscript $IR$ stands for derived from IR-selected sources) by convolving the volume density
of the proxy (L$_{IR}$) with the M$_{dust}$-L$_{IR}$ relation as:

\begin{equation}
 \rho_{dust,IR}=\int{ <M_{dust}> \phi(x) dx}
 \end{equation}

\noindent where now $x$ is L$_{IR}$ and $\phi(x)$ is the IR-luminosity
function (IRLF). As done for L$_{UV}$, for converting the mean
relation  from log space to linear space we have taken into account the
dispersion, although given the small dispersion (${\lsimeq}0.05$ dex), the
second term in Eq. \ref{eq2} is negligible.

As for the $\rho_{dust,UV}$, also for $\rho_{dust,IR}$ we consider two
quantities: a lower limit obtained
integrating the IRLF only in the
 IR luminosity range sampled by the data ($11.7<$log(L$_{IR}$)$<13$), and the
 value ($\rho_{dust,IR}$) obtained integrating the IRLF in the commonly used IR
range ($\rho_{dust,IR}$; $9<$log$(L_{IR})<14$, see
\citealt{2020A&A...643A...8G}). 

The uncertainty on $\rho_{dust,IR}$ depends on the
uncertainties affecting  the IRLF and on the uncertainties affecting the log(M$_{dust}$)-log(L$_{IR}$)
relation. The former has been
computed considering the 1$\sigma$ uncertainty derived through the
MCMC analysis of the IRLF estimated by \cite{2020A&A...643A...8G} (see their Table 4). This term accounts for a 1$\sigma$ dispersion of $\sim$0.3-0.5 dex in
$\rho_{dust,IR}$, significantly higher than that derived from the
dispersion of the log(M$_{dust}$)-log(L$_{IR}$) relation ($\lsimeq0.05$ dex).

The results are shown in Fig. \ref{dmd_fig} and reported in Table
\ref{table3}.



\begin{table}
\centering
  \begin{tabular}{c|c|c|c} \hline\hline
    redshift    & $N$    &   log$L_{UV} $     &     log$M_{dust}$  \\
                   &                     &   [$L_{\odot}$]          &           [$M_{\odot}$]  \\\hline
    $4.4 < z < 4.6$     &   23 (2) & 10.4 - 10.9      & 7.53$^{+0.15}_{-0.17}$     \\    
                  &  23 (8)              & 10.9 - 11.1       &   7.83$^{+0.13}_{-0.14}$\\
                  &   21 (5)               & 11.1 - 11.6       &   7.83$^{+0.11}_{-0.14}$\\\hline
    $5.1 < z < 5.9$     &   25 (0)               & 10.4 - 10.9        &     $<$7.5 \\
                 &  10 (1)                & 10.9 - 11.1       &    $<$7.6 \\
                 &   15 (7)                & 11.1 - 11.6       &  7.72$^{+0.2}_{-0.2}$ \\\hline\hline                                                                              \end{tabular}
    \caption{Results of the stacking analysis. Median M$_{dust}$ are
      reported, derived from the median stacked fluxes obtained as
      described in Sect. \ref{dmd_uv_sec} and using equation \ref{eq_mdust} as for the
      targets. The upper limits are at 3$\sigma$. The number of
      sources stacked in each bin and indivially detected are also reported.  }
  \label{table2}
\end{table}

\subsection{Discussion on the DMD derivation}
\label{dmd_fir_sec}

In Fig. \ref{dmd_fig} we report the evolution of the comoving DMD as a function of the look-back time from our analysis together
with determinations from
other observational works and with model predictions. The ALPINE
derivations are reported at 12~$<t_{lookback}<~12.5$ Gyrs, corresponding
to the redshift range of the ALPINE sources (in case of the derivation
from the UV-populations a slightly wider range is considered for graphic convenience).
The values obtained from the UV-selected population (IR-selected population), obtained by integrating the
UVLF (IRLF) in the extended UV (IR) luminosity range are represented
by a blue (red) point. The boxes around the points represent the 1$\sigma$ associated
uncertainties, with the large dispersion of ${\rho_{dust,UV}}$ reflecting the large dispersion in the
log(M$_{dust}$)-log(L$_{UV}$) relation (see Fig. \ref{mdust_versus_luv_fig}, Top).

The value of ${\rho_{dust,UV}}$ is $\sim$30 \% of ${\rho_{dust,IR}}$ (although consistent within 1$\sigma$), and this is not surprising, since the
    UV-selected population is not representative, by construction, of the most dusty obscured
    galaxies which are present in the blind IR
    selection. This is supported by the comparison of the L$_{IR}$
    distribution of the 
    ALPINE targets and of the serendipitously detected sources. The serendipitous sources
    show $<$log(L$_{IR})$>$=12.1^{+0.9}_{-0.3}$, hence they are  
    generally brighter than the ALPINE targets, with $<$log(L$_{IR})>=11.6^{+0.1}_{-0.2}$ 
    (see Fig. \ref{mdust_versus_lir_fig}, Botton). While this difference alone cannot 
    be conclusive since it could be partially caused by different S/N cuts (S/N=3.5 and 5, for our targets and the
    serendipitously sources, respectively, see \citealt{2020A&A...643A...2B}), beside 
    their different IR luminosity distribution, also the 
    fact that only half of the serendipitous sources (see \citealt{2020A&A...643A...8G}) are detected
    in the UV  at magnitudes 
    $z{\sim}{25.9}$ (\citealt{2016ApJS..224...24L}) further supports our hypothesis. 

Our result is also in line with the one found recently by 
\cite{2020A&A...643A...8G} which concerns the star-formation rate
density (SFRD). By deriving the SFRD from IR sources serendipitously detected in ALPINE, the authors found that the difference with literature UV results
increases with redshift, reaching a factor $\sim$10 at $z{\sim}6$.

In Fig. \ref{dmd_fig} we also show previous determinations of
$\rho_{dust}$ (see references in the figure legend and caption). 
In the redshift range sampled by the ALPINE survey, recent determinations were derived by \cite{2020ARA&A..58..363P} and  
  \cite{2020ApJ...892...66M}.  \cite{2020ARA&A..58..363P} (salmon filled squares in Fig. \ref{dmd_fig}) estimated the DMD indirectly, by
  convolving the mass density of neutral gas with the dust-to-gas
  (DTG) ratio. The density of neutral gas was derived from an
  extensive literature collection of different
  measurements,  mainly based on high-$z$ quasar spectra absorbing
  features, whereas the DTG was estimated from the depletions of
  different heavy elements into the solid
  phase. The determinations from \cite{2020ARA&A..58..363P} are consistent with our  $\rho_{dust}$ value obtained from 
  UV-selected galaxies, whereas it is lower than our
  `fiducial value' derived from IR selected
  galaxies. However, \cite{2020ARA&A..58..363P} cautioned the reader against a
  potential selection bias affecting their estimate (see their Sec. 3.2.3), as extremely dusty systems might be missing from
  their optically-selected quasar sample.   The large uncertainty (up to 50$\%$) affecting our
  determination of $\rho_{dust}$ for IR selected galaxies reflects the small sample used to estimate
  the IRLF. This stresses further the need for future investigations 
  and for future space IR high-sensitivity instruments suitable for tracing the dust content of high-redshift
  galaxies.

 \cite{2020ApJ...892...66M} estimate the DMD directly, thanks
  to ALMA observations from the ASPECS LP survey (yellow filled squared in figure). The ASPECS LP survey (\citealt{2020ApJ...897...91G}) was
obtained in Band 6 (1.2 mm), over an area of ${\sim}{4.2}$
arcmin$^{2}$. The reported determinations from
the ASPECS survey from $0.3<z<5.5$ have been computed by stacking the
ALMA maps at the positions of $H$-band selected galaxies above a stellar mass
of 10$^8$ M$_{\star}$ in distinct redshift bins. The points
corresponding to $3.2<z<4.5$ ($4.5<z<5.5$), have been obtained by
stacking 44 (9) galaxies, but, since the sample can not be
considered as stellar mass complete, these measurements  can only be
considered lower limits (see Table 1 in
\citealt{2020ApJ...892...66M}). Our values of $\rho_{dust}$ obtained from the UV-selected and the FIR-selected
galaxies, at the redshift of the ALPINE survey (4.5$<z<$5.9), are 
consistent with the lower limits from the ASPECS survey
obtained in the two highest redshift bins ($3.2<z<4.5$ and $4.5<z<5.5$).

The general DMD as a function of the cosmic
time shows a mild increase from $z{\sim}$5
up to the cosmic noon ($1<z<3$) (a factor of $\sim$2.5 in our analysis),
followed by a smooth decline (a factor of ${\sim}$3) up to the local
Universe (see also \citealt{2018MNRAS.475.2891D}, \citealt{2020MNRAS.491.5073P}, \citealt{2020ApJ...892...66M}).

By comparing the DMD and the SFRD, as already stated by \cite{2020ApJ...892...66M}, at
$z<2$ both quantities show a decline toward the local Universe, but 
the decline of the DMD is less pronounced than
that observed for the SFRD (a factor $\sim$3-4 instead of $\sim$8). At $z>3$,
the SFRD derived from IR data shows an almost flat behaviour (\citealt{2020A&A...643A...8G}
and references within), while the DMD shows a mild decline (a factor $\sim$2.5). We postpone the
interpretation of these data to a theoretical work. Nevertheless, we 
stress that the cosmic dust mass at $z{\sim}$5 may not have reached its maximum value yet,  
and this could be partially explained considering the typical dust production timescales, i.e. 
the time after which the dust mass reaches its maximum in galaxies, 
which ranges from a few 0.1 Gyr up to several Gys in spirals and proto-spheroids, 
respectively (see panel a) in Fig. \ref{scaling_fig}). 

In Fig. \ref{dmd_fig} we also show the predictions from models. 
The model from \cite{2017MNRAS.471.4615G} is the one that best 
reproduces the global trend shown by  the DMD evolution 
from the local Universe up to $z{\sim}5$. This is not a cosmological model, but a phenomenological
one that combines the chemical evolution of different galaxy types
(\citealt{2008A&A...479..669C}) with the evolution of 
different galaxies as derived from
\cite{2015ApJ...803...35P}, based on the IRLF from
\cite{2013MNRAS.432...23G}. The results of \cite{2017MNRAS.471.3152P} and
\cite{2019MNRAS.489.4072V} were obtained by means of cosmological, semi-analytical galaxy formation  models. 
They both roughly account for the observed decrease of the DMD at high-z ($z>3$), but they
fail to reproduce the decline at lower $z$ and they overestimate
the dust mass budget detected locally in resolved galaxies. 

On the other hand, the models from \cite{2018MNRAS.478.4905A} and
 \cite{2019MNRAS.490.1425L} are based on cosmological hydrodynamic
 simulations. They are 
 globally inconsistent with the data, severely overestimating the DMD at
 $z<1$. At higher redshift, the model from
 \cite{2018MNRAS.478.4905A} is in fair agreement with the
 observations, whereas the model from \cite{2019MNRAS.490.1425L}
 underestimates our 'fiducial' DMD by  up to a factor of $\sim$10.

Clearly, the DMD at high-$z$ needs further
investigation from both an observational and a theoretical point of
view. From the observational side, we stress that the DMD is estimated assuming our 'fiducial' temperature of T=25 K, a choice made also
  in several other studies (e. g. \citealt{2014ApJ...783...84S}, \citealt{2020ApJ...892...66M}).
  As stated in Sec. \ref{dust_sec}, a warmer dust mass (T= 35 K) would
  reduce the DMDs by a factor of 60\%.

  In order to shed more light
on the trend of the DMD at z$>$2, in particular in relation with the
cosmic SFRD, a significant step forward will be achieved from the
synergy of far-IR and sub-mm/mm facilities, allowing to consider at least two dust
components, i.e. the warm and the cold one. This will be possible when the ALMA
data will be complemented by data from a future far-IR satellite similar to SPICA, with a sensitivity more
than an order of magnitude better than Herschel
(\citealt{2018PASA...35...30R}). 

\begin{table}
\centering
  \begin{tabular}{c|c} \hline\hline
 \multicolumn{2}{c}{log($\rho_{dust}$)    z${\sim}5$ } \\\hline

log($\rho_{dust,UV}$)                    &   $log(\rho_{dust,IR}$) \\
\scriptsize [M$_{\odot}$Mpc$^{-3}$]       & \scriptsize[M$_{\odot}$Mpc$^{-3}$]               \\\hline
                                       & \\
      4.8$^{+0.4}_{-1.0}$            &     5.3$^{+0.5}_{-0.3}$ \\
                                       & \\
        $>4.3$                       & $>4.8$        \\\hline\hline
        \end{tabular}

    \caption{Cosmic dust mass density derived from the ALPINE survey
      at $z{\sim}{5}$. The left column reports the measurement derived
      from the ALPINE UV-selected galaxies using L$_{UV}$ as proxy for
      M$_{dust}$ and integrating the UVLF in the typical
    $0.03L^{\star}-100L^{\star}$ range. The right column reports the 
      measurement derived from the serendipitous IR-detected sources using L$_{IR}$ as proxy for
      M$_{dust}$ and integrating the IRLF in the typical
      $9<log(L_{IR})<14$ range. We also report in the second row the
      lower limits obtained by integrating the UVLF and IRLF only in the
      range sampled by the detected sources ($10.4<$log$(L_{UV})<11.5$ and
      $11.2<$log$(L_{IR})<13$, respectively). }
  \label{table3}
\end{table}
\normalsize

\section{Summary}
\label{summary_sec}
We use observations of the ALPINE survey to study the dust
mass content of normal star-forming galaxies at high-$z$ ($z{\sim}$5)
and to provide, for the first time, an estimate of the cosmic dust
 mass density (DMD) at such high look-back time.

ALPINE is a targeted survey specifically designed to detect the bright [CII]
158~$\mu$m line in 118 UV-selected galaxies in the redshift range
(4.4$<z<$5.9). For the aims of the present analysis, we consider the
rest-frame FIR continuum emission of the ALPINE galaxies (individually
detected or their stacks) and the FIR continuum emission of the blind
serendipitously detected galaxies in the
ALPINE area.

Dust masses are measured from the continuum
emission at 250 ${\mu}$m (see \citealt{2014A&A...562A..67G}), extrapolated from the 157 $\mu$m ALMA data, 
assuming a mass-weighted temperature of 25 K and a dust emissivity of $\beta$=1.8 
 Our main results can be summarized as follows:

\begin{itemize}
\item{We combine the dust masses of the 23 ALPINE continuum detected
    sources with other robustly determined physical parameters derived from SED-fitting
    (i.e. Age, M$_{\star}$, SFR) or from the ALPINE observations themselves
  (i.e. M$_{gas}$) to settle the evolutionary stage of normal
  star-forming galaxies at $z{\sim}$5. To this purpose, we compare the
  observed dust scaling relation with the evolutionary tracks
  predicted by chemical models for spiral galaxies and for precursors of local elliptical galaxies
  of different masses (called proto-spheroids, PSPH).
  From the analysis of several scaling relations, we 
  conclude that our galaxies show dust masses and SFR
  values typically
  consistent with intermediate- and low-mass PSPH models.
  Our models indicate that galaxy discs, such as that of the MW,
  show at early epochs SFR values $<10~M_{\odot}/yr$ and dust masses much lower
  than the ones measured in ALPINE galaxies. 
   Our conclusion is confirmed  also by the non-detected
   galaxies, which, even if less extreme than the actual detections,
   in the SFR-Age diagram generally show SFR values compatible
   with the ones of the massive discs which, in the galactic downsizing picture,
   at these epochs are expected to be more intensely star-forming than  
   low- and intermediate- mass spirals.
   It is worth noting that our results do not exclude that the 
   ALPINE galaxies might be compatible 
   with the progenitors of MW-like galaxies, whose earliest assembly and star-formation activity 
   is expected to occur mostly in their spheroidal components, i.e. in their bulges (e. g.,
   \citealt{2012MNRAS.427.1401C}; \citealt{2016ApJ...824...94R}; \citealt{2020MNRAS.494.5936F}). }

 \item{We estimate $\rho_{dust}$ at $z{\sim}$5 for UV-selected and
    FIR-selected galaxies, using L$_{UV}$ and L$_{IR}$ as proxies for M$_{dust}$, respectively. In the first case, 
    we use the log(M$_{dust}$)-log(L$_{UV}$) relation found for the
    ALPINE galaxies (considering both the detected
    and the non-detected sources) and the convolution with the UVLF; in the second case, we consider the
    log(M$_{dust}$)-log(L$_{IR}$) relation found for the 10 blindly FIR-detected
    sources and the convolution with the IRLF. The derived
    $\rho_{dust, UV}$ is $\sim$30 $\%$ $\rho_{dust,IR}$, although the two estimates are
    marginally consistent at 1$\sigma$. Our result supports the
    conclusion that UV-selected galaxies miss the most obscured/dusty objects and we consider
    $\rho_{dust,IR}$ as our fiducial value.}
    
\item{We compare our fiducial $\rho_{dust}$ at $z{\sim}$5 with the
    predictions from models, taking into account also the $\rho_{dust}$
    determinations at lower $z$ from the literature. The phenomenological model from \cite{2017MNRAS.471.4615G}
    is the model that best 
    reproduces the observed evolution of $\rho_{dust}$.
    Since this model is basically built upon
    the observed evolution of the cosmic SFR, this result indicates that it can
    roughly account for the expected timescales for dust production in galaxies. 
    On the other hand, the ab-initio cosmological, semi-analytical
    models of \cite{2017MNRAS.471.3152P} and \cite{2019MNRAS.489.4072V} and 
    and the  cosmological simulations of \cite{2019MNRAS.490.1425L}
    roughly account for the observed evolution of $\rho_{dust}$ dust at $z>1$,
    but they overestimate the dust mass budget at lower redshift, failing in reproducing the decreasing
    trend observed at $z<1$. These results indicate that our physical understanding the cosmic evolution of the
    dust mass needs to be improved. }
  
Our study strongly outlines the need for statistical samples of
 galaxies at high-$z$ with a full characterisation of the IR
 spectrum,  from the Wien up to the Rayleigh-Jeans regimes. This can be achieved by exploiting the synergies between sub-mm/mm (i.e. ALMA, NOEMA) and far-IR facilities.

\end{itemize}

\begin{figure*}
\includegraphics[width=0.8\textwidth,angle=0]{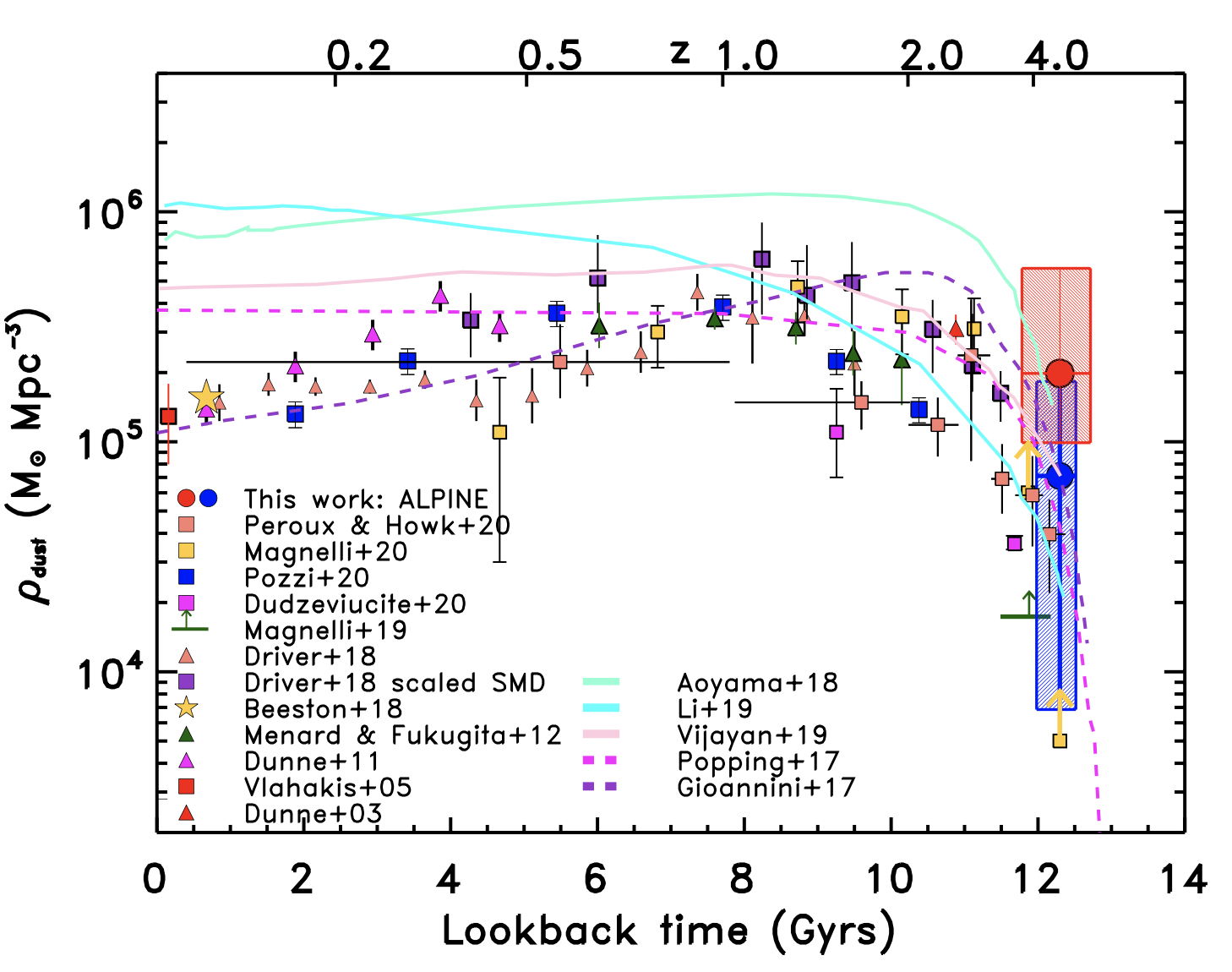}
\caption{Evolution of the comoving dust mass density
  ($\rho_{dust}$) as a function of the look-back time. Our
  determinations are shown at
  $t_{lookback}{\sim}12.3$ Gyrs,
obtained from the ALMA continuum detection of the
  UV-selected (blue circle) and serendipitously detected sources (red circle), respectively. The
  blue/red boxes represent the 1$\sigma$ uncertainties. For
  comparison, estimates from other surveys are shown (ALMA:
  \citealt{2020ApJ...892...66M}; Herschel:
  \citealt{2011MNRAS.417.1510D}, \citealt{2018MNRAS.475.2891D}, \citealt{2018MNRAS.479.1077B}, \citealt{2020MNRAS.491.5073P}; SCUBA:
  \cite{2003MNRAS.341..589D}, \citealt{2005MNRAS.364.1253V}, \citealt{2020arXiv201006605D}; IRAM: \citealt{2019ApJ...877...45M};  absorbers in the optical spectra of quasars:
  \citealt{2012ApJ...754..116M}, \citealt{2020ARA&A..58..363P}). The lines represent
  different models: the cosmological hydrodynamic
  models from \cite{2018MNRAS.478.4905A} and
  \cite{2019MNRAS.490.1425L} as water-green and light blue lines,
  respectively; the semi-analytical models from \cite{2017MNRAS.471.3152P} and
\cite{2019MNRAS.489.4072V} as pink and dashed-magenta lines,
respectively; the chemical evolution model from
\cite{2017MNRAS.471.4615G} as dark-green dashed line.}
\label{dmd_fig}
\end{figure*}

\begin{acknowledgements}
This paper is dedicated to the memory of Olivier Le F\`evre, PI of the 
ALPINE survey. We are grateful to an anonymous referee for valuable
suggestions that improved the paper. FP thanks Simone Bianchi for useful discussion on the dust
properties. We acknowledge support from grant PRIN MIUR
2017- 20173ML3WW$\_$001. FP also acknowledges support from the INAF
man-stream program 'Gas-DustPedia: A definitive view of the ISM in the Local Universe'.
FC also acknowledges support from the INAF main-stream (1.05.01.86.31). D.R. acknowledges support from the National Science Foundation  under grant numbers AST-1614213 and AST-1910107. D.R. also
   acknowledges support from the Alexander von Humboldt Foundation
   through a Humboldt Research Fellowship for Experienced
   Researchers. G.C.J and R.M  acknowledge ERC advanced grant 695671
   'QUENCH' and support by the Science and Technology Facilities
   Council (STFC). 
\end{acknowledgements}

\bibliographystyle{aa}
\bibliography{pozzi}

\begin{appendix}
\section{Dust mass estimates}

\begin{table}[b]
  \centering
\caption{M$_{dust}$ estimates for the 23 ALPINE continuum detected
    sources. The dust masses have estimated using a Modified Black
    Body spectrum (MBB), under the approximation of thin emission, assuming
    T$_{dust}$=25 K and the spectral index $\beta=1.8$ (see
    Sec. \ref{dust_sec}). $^{a}$ Multi-component objects (see Appendix D.2
and Table D.1 in \citealt{2020A&A...643A...2B}). For
DEIMOS$\_$COSMOS$\_$881725 the sum of the components have been
considered; for vuds$\_$cosmos$\_$51012097 and
vuds$\_$efdcs$\_$530029038 only the fluxes of the central targets
since the companions are likely separated objects (see
Sec. \ref{data_sec}).}.
 \begin{tabular}{l|l} \hline\hline
    Target Name   & log M$_{dust}$  \\
                 &   [log$_{10}$(M/M$_{\odot}$)]                      \\\hline
      CANDELS$\_$GOODSS$\_$19          &    8.21$\pm$0.19     \\
      CANDELS$\_$GOODSS$\_$32          &    8.12$\pm$   0.21      \\
   DEIMOS$\_$COSMOS$\_$396844          &    8.31$\pm$   0.19    \\
   DEIMOS$\_$COSMOS$\_$417567          &    8.22$\pm$   0.21      \\
   DEIMOS$\_$COSMOS$\_$422677          &    8.33$\pm$   0.22     \\
   DEIMOS$\_$COSMOS$\_$460378          &    7.95$\pm$   0.21      \\
   DEIMOS$\_$COSMOS$\_$488399          &    8.32 $\pm$ 0.18     \\
   DEIMOS$\_$COSMOS$\_$493583          &    8.14 $\pm$ 0.22      \\
   DEIMOS$\_$COSMOS$\_$494057          &    8.16 $\pm$  0.18      \\
   DEIMOS$\_$COSMOS$\_$539609          &    8.13 $\pm$  0.21      \\
   DEIMOS$\_$COSMOS$\_$552206          &    8.35 $\pm$  0.20       \\
    DEIMOS$\_$COSMOS$\_$683613         &    8.29$\pm$   0.19     \\
   DEIMOS$\_$COSMOS$\_$818760          &    8.81$\pm$   0.17      \\
   DEIMOS$\_$COSMOS$\_$848185          &    8.38$\pm$   0.18       \\
   DEIMOS$\_$COSMOS$\_$873756          &    8.91$\pm$   0.17       \\
   DEIMOS$\_$COSMOS$\_$881725          &    8.32$\pm$0.25$^{a}$        \\
   vuds$\_$cosmos$\_$5100822662        &    8.09$\pm$   0.18      \\
   vuds$\_$cosmos$\_$5100969402        &    8.29$\pm$   0.21      \\
   vuds$\_$cosmos$\_$5100994794        &    7.85$\pm$   0.21      \\
   vuds$\_$cosmos$\_$5101209780        &    8.27$\pm$ 0.23$^{a}$       \\
   vuds$\_$cosmos$\_$5101218326        &    8.44$\pm$  0.18   \\
   vuds$\_$cosmos$\_$5180966608        &    8.39$\pm$  0.19     \\
   vuds$\_$efdcs$\_$530029038          &    7.85$\pm$ 0.26$^{a}$       \\\hline
  \end{tabular}
\label{table1}
\end{table}

\end{appendix}

\end{document}